\documentclass[%
 reprint,
 amsmath,amssymb,
 aps,
]{revtex4-2}

\usepackage{graphicx}
\usepackage{dcolumn}
\usepackage{bm}
\usepackage{physics}

\usepackage[utf8]{inputenc}
\usepackage[english]{babel}
\usepackage[T1]{fontenc}
\usepackage{amsmath}
\usepackage{hyperref}

\usepackage{tikz}
\usepackage{lipsum}
\usepackage{xcolor}
\usepackage[ruled]{algorithm2e}

\usepackage{graphicx}
\usepackage{dcolumn}
\usepackage{bm}

\usepackage{amsmath}
\usepackage{accents}
\newlength{\dhatheight}
\newcommand{\doublehat}[1]{%
    \settoheight{\dhatheight}{\ensuremath{\hat{#1}}}%
    \addtolength{\dhatheight}{-0.35ex}%
    \hat{\vphantom{\rule{1pt}{\dhatheight}}%
    \smash{\hat{#1}}}}

\begin{document}

\preprint{APS/123-QED}

\title{Deep Quantum Neural Networks are Gaussian Process}


\author{Ali Rad \\
\textit{Joint Center for Quantum Information and Computer Science (QuICS)}\\
 \textit{Joint Quantum Institute (JQI),
  University of Maryland, College Park, Maryland 20742, USA}\\
 \texttt{rad@umd.edu}
 }

\date{\today}
\begin{abstract}
    The overparameterization of variational quantum circuits, as a model of Quantum Neural Networks (QNN), not only improves their trainability but also serves as a method for evaluating the  property of a given ansatz by investigating their kernel behavior in this regime. In this study, we shift our perspective from the traditional viewpoint of training in parameter space into function space by employing the Bayesian inference in the Reproducing Kernel Hilbert Space (RKHS). We observe the influence of initializing parameters using random Haar  distribution results in the QNN behaving similarly to a Gaussian Process (QNN-GP) at wide width or, empirically, at a deep depth. This outcome aligns with the behaviors observed in classical neural networks under similar circumstances with Gaussian initialization.   Moreover, we present a framework to examine the impact of finite width in the closed-form relationship using a $ 1/d$ expansion, where $d$ represents  the dimension of the circuit's Hilbert space. The deviation from Gaussian output can be monitored by introducing new quantum meta-kernels.
    Furthermore, we elucidate the relationship between GP and its parameter space equivalent, characterized by the Quantum Neural Tangent Kernels (QNTK)
. This study offers a systematic way to study QNN behavior in over- and under-parameterized scenarios, based on the  perturbation method, and  addresses the limitations of tracking the gradient descent methods for higher-order corrections like dQNTK and ddQNTK. Additionally, this probabilistic viewpoint lends itself naturally to accommodating noise within our model.

\end{abstract}

\maketitle

\section{Introduction}
The capacity for learning and trainability in quantum circuits is an intriguing subject that merits further exploration and discovery. The issue of how trainable variational quantum circuits and quantum neural networks (QNNs) poses a significant challenge that needs to be tackled. The standard approach to training  QNN is within the context of \textit{parameter space}. The aim is to find the optimum point in the loss landscape using local gradient-based methods. However, these methods might encounter the Barren Plateau problem \cite{mcclean2018barren}, where the magnitude of the gradient decreases exponentially as the size of the Hilbert space increases.

Alternatively, we can shift the training process into \textit{functios space}, viewing QNNs as linear combinations of Kernels that exist within the Reproducing Kernel Hilbert Space (RKHS)\cite{schuld2021supervised,schuld2019quantum}. The learning task in the parametric space of gate's parameters can be transformed into the task of identifying the correct coefficients of kernel regression parameters, which are of the order of the input dataset. This is typically much less than the dimension of the Hilbert space. The process of identifying these coefficients is usually more convex than implementing the original gradient descent method in the parameter space.

Although the quantum kernel's behavior in different quantum circuit regimes is still a largely uncharted area, recent findings suggest that kernel-based methods can be more successful than gradient methods in mitigating the Barren Plateau problem in under-parameterized regimes\cite{rad2022surviving,duffield2023bayesian}. With this evidence, we can now turn our attention in this work to the other part of the spectrum: the over-parameterized regime. We hope that this will enhance our understanding and lay the groundwork for more generic and practical applications in the NISQ era.
In this regime, both the functional and parametric methods present more straightforward and comprehensible approaches. Similar to classical Neural Networks (CNNs)\cite{jacot2018neural}, the output of the QNN can be determined using the kernel regression method. The kernel employed in this lazy training method is referred to as the Quantum Neural Tangent Kernel (QNTK)\cite{liu2023analytic,shirai2021quantum,liu2022representation,liu2022laziness,abedi2023quantum}.

An interesting observation has been made regarding CNN when they are over-parameterized through wide layers or deep networks: their output tends to converge towards a Gaussian distribution and Gaussian Process (GP) \cite{Neal1996PriorsFI,lee2017deep,novak2018bayesian,lee2019wide}. Given this observation, it's feasible to replace the prior distribution of parameters in classical neural networks with a Gaussian process prior. This process shifts the focus from parameter spaces to the space of functions. A Gaussian process is a probabilistic model that defines a distribution over functions. the idea is to use a prior probability on functions, which is specified as a Gaussian process that determines the shape of the function space, and then update this with the observed data to get a posterior distribution over functions. In contrast to kernel regression, in which one  typically needs to make decisions about the kernel hyperparameters, while in Gaussian process regression, these are typically learned from the data.
Furthermore, while kernel regression provides a point estimate, Gaussian process regression provides a full probabilistic model, allowing for uncertainty estimation in predictions. we can think of Gaussian process regression as a probabilistic, kernel-based method, and it's a generalization of kernel regression that also provides uncertainty estimates.

\begin{figure*}\label{Fig:Feature_sapce}
  \centering
  \includegraphics[width=\textwidth]{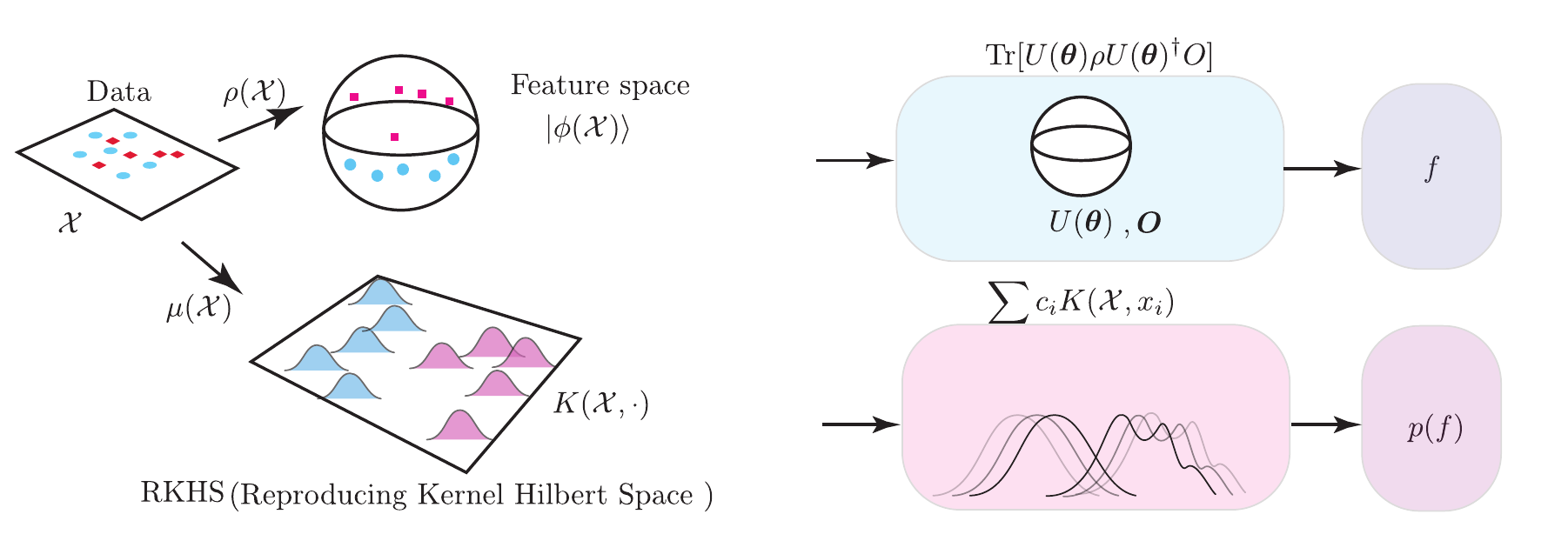}
  \caption{ Encoding the classical or quantum data can be  seen in two way. (top): Encoding in 
Feature space which is  Hilbert space, $|\phi(x)\rangle  \in F$. The quantum circuit evolves the feature vector using a parameter $\theta$ to predict the label. Training aims to find the optimal $\theta$ that achieves the highest criteria, such as minimizing the loss function, for accurate label prediction.
({button}):Encoding data as a  quantum kernel $K(\mathcal{X},\cdot)$. One example of quantum kernel is Fidelity kernel defines as: $K^F(x,x')=|\langle \phi(x),\phi(x')\rangle|^2=\Tr[\rho(x) \rho(x')]$. The Reproducing Kernel Hilbert Space (RKHS) is the  feature space that has canonical feature space if for $K(\cdot, x) \in \mathcal{H}$ for all $x\in \mathcal{X}$ it has a reproducing property $f(x)=\langle f,K(\cdot, x)\rangle_\mathcal{H}$. In this method, a training method for quantum circuits that seeks to find the optimal coefficients $\{c_i\}$, which effectively describe the output function $f(x) = \sum_{i} c_i K(x, x_i)$ based on the observed data points $\{x_i\}$. This technique is commonly referred to as kernel regression. This method can be extended to Bayesian inference which probabilistic kernel regression and predict the $p(f)$ instead of $f$.}
\end{figure*}

\subsection{Quantum Meta Kernels}


We begin by study the standard structure of variational quantum circuits, integral to quantum neural network models. The structure is given by:
\begin{eqnarray}\label{Eq:framework}
U(\boldsymbol{\theta})=\prod_{\ell=1}^L U_i(\theta_i)W_i,
\end{eqnarray}
where, $U_i(\theta_i)$ represents a unitary gate with a variational parameter $\theta_i$. $W$ stands for a fixed and unparameterized section of the circuit. In a common scenario, $U_i(\theta_i)$ can be represented as $U_i(\theta)=e^{i\theta_i X_i}$, where $X_i$ is a Hermitian operator. The layer depth is denoted by $L$, which is equivalent to the number of parameters $\boldsymbol{\theta}=\{\theta_1, \cdots, \theta_L\}$.
By defining $\tilde{U}_{i}(\theta_i)(\rho):=U_i(\theta_i)\rho U_i^\dagger(\theta_i)$ ans $\tilde{W}_i:=W_i \rho W_i^\dagger$.
We can express the whole quantum circuit as a quantum model that for a given input $\rho_\alpha$, it return an output (array or scalar):
\begin{eqnarray}
f_{i,\alpha}=\Tr(\mathcal{U}(\boldsymbol{\theta})(\rho_\alpha)O_i)
\end{eqnarray}
such that 
\begin{eqnarray}\label{UO}
    \mathcal{U}(\boldsymbol{\theta})(\cdot)=\bigcirc_{\ell=1}^L (\tilde{U}_{\ell}(\theta_\ell)\circ \tilde{W}_\ell)(\cdot)
\end{eqnarray}

In order to analyze the probabilistic behavior of the outputs from our quantum models, it's necessary to examine the k-point correlation function, given by

\begin{equation}
\mathbb{E}_{\boldsymbol{\theta}}[f_{i_1;\alpha_{1}}(\boldsymbol{\theta})\cdots f_{i_k;\alpha_{k}}(\boldsymbol{\theta})]=\mathbb{E}_{\boldsymbol{\theta}}[\Tr(\mathcal{U}^{(k)}({\boldsymbol{\theta}})(\rho_\alpha^{\otimes k })O^{\otimes k})]
\end{equation}

Where $\mathcal{U}^{(k)}$ represents an extension of Eq.\ref{UO}, which is defined as follows:
\begin{equation}
\mathcal{U}^{(k)}({\boldsymbol{\theta}})(\rho_\alpha^{\otimes k }):=\bigcirc_{\ell=1}^L (\tilde{U}^{\otimes k}_{\ell}(\theta_\ell)\circ \tilde{W}^{\otimes k}_\ell)(\rho_\alpha^{\otimes k })
\end{equation}

If the gates in a quantum model satisfy the t-design condition
\begin{equation}
    \frac{1}{|S|}\sum_{s=1}^{s=|S|}P_{t,t}(U)=\int_{\tilde{U}}d\mu(U) U P_{t,t}(U)
\end{equation}
then, it is possible to analytically evaluate the expectation values. This holds true for overparameterized and deep circuits, where evidence suggests that the $t$-design condition can be met. In such cases, we can derive analytical and closed-form expressions for these expectations for a random Haar measure ensembles: 
\begin{align}\label{Eq:Haar_average}
    \begin{split}
        &\int_{\tilde{U}(d)} U_{i_1,j_1}\cdots U_{i_p,j_p}U^\dagger_{i',j'}\cdots U^{\dagger}_{i'_p,j'_p}d\mu(U)\\
        &=\sum_{\alpha,\beta\in S_p}\delta_{i_1 i'_{\alpha(1)}}\cdots \delta_{i_p i'_{\alpha(p)}} \delta_{j_1 j'_{\beta(1)}}\cdots \delta_{j_p j'_{\beta(p)}}W_{g,d}(\alpha^{-1}\beta),
    \end{split}
\end{align}
where The $W_{g,d}$ is Weingarten function defined on  the symmetric group $S_p$ \cite{fukuda2019rtni,puchala2011symbolic}

\begin{figure*}\label{Fig:Full_gaussian}
  \centering
  \includegraphics[width=\textwidth]{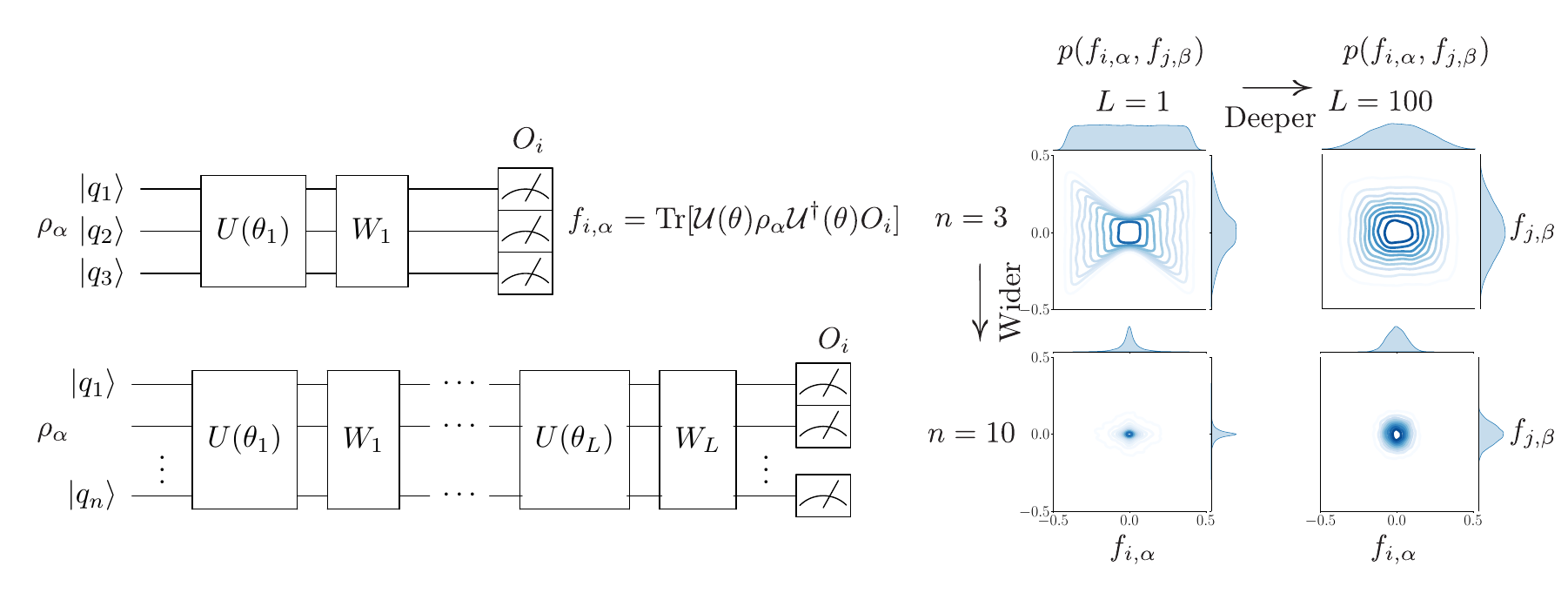}
  \caption{ In variational quantum circuits, as a  model for quantum neural networks (QNNs), the number of qubits, denoted as $n$ and defined as $n=\log(d)$, can be interpreted as the width of the model. To investigate the impact of the width ($n$) and depth ($L$) on the output distribution, we conducted simulations of a quantum circuit following the described structure. Each layer of the circuit consists of random two-qubit gates followed by entanglement gates. The observables used in the simulations were constructed from the Pauli $Z$ matrix, forming a complete basis. We generated $10^{5}$ random Haar instances for each circuit with given values of $n$ and $L$. The joint distribution, $P(f_i, f_j)$, of the output was plotted for two fixed indices $i$ and $j$. The input of the circuit was constructed from the MNIST dataset, reduced size with the Principal Component Analysis(PAC) , and then mapped to the feature space using the ZZFeaturemap\cite{havlivcek2019supervised}. As observed, as the quantum neural network (QNN) becomes deeper, the output distribution tends to converge more towards a Gaussian distribution. }
\end{figure*}

To gain a deeper understanding of the non-Gaussian characteristics of the distribution, it is often beneficial to study the connected $n$-point correlator. This can be achieved by removing the contributions from the \textit{vacuum} state in our observable, according to the  Wick's theorem:
\begin{align}
    \begin{split}
&\mathbb{E}_{\boldsymbol{\theta}}[f_1 f_2 \cdots f_n]=\mathbb{E}_{\boldsymbol{\theta}}[f_1 f_2 \cdots f_n]|_{\text{connected}}\\
&+\sum_{\text{all pairing}}\mathbb{E}_{\boldsymbol{\theta}}[f_{\alpha_{1,1}} \cdots f_{\alpha_{m_1,1}}] \cdots \mathbb{E}_{\boldsymbol{\theta}}[f_{\alpha_{m,1}} \cdots f_{\alpha_{m,m}}],
    \end{split}
\end{align}
where the term "all-pairing" denotes all feasible subgroups with a collective count of $m$. By adjusting the $f \rightarrow f-\mathbb{E}[f]$ we can observe that  $\mathbb{E}[f]|{\text{connected}}=\mathbb{E}[-f]{\text{connected}}$ This assumption implies that the odd order of correlators will vanish.



Through careful calculation, as described in the appendix, we derive the following result for two-point correlator : 
\begin{align}
    \begin{split}
        &\mathbb{E}[f_{i_1,\alpha_1}f_{i_2,\alpha_2}]|_{\text{connected}}\\
    &=\frac{\text{Tr}\left(O_{i_2}O_{i_1}\right)}{d-d^3}+\frac{\text{Tr}\left(O_{i_2}O_{i_1}\right)\text{Tr}\left(\rho_{\alpha_1}\rho_{\alpha_2}\right)}{d^2-1}\\&+\frac{\text{Tr}\left(O_{i_1}\right)\text{Tr}\left(O_{i_2}\right)}{d^2-1}+\frac{\text{Tr}\left(O_{i_1}\right)\text{Tr}\left(O_{i_2}\right) \text{Tr}\left(\rho _{\alpha_1}\right) \text{Tr}\left(\rho_{\alpha_2}\right)}{d^2}
    \end{split}
\end{align}
and similarly, for four-point correlator we obtain:
\begin{align}
    \begin{split}
        &\mathbb{E}[f_{i_1,\alpha_1} f_{i_2,\alpha_2} f_{i_3,\alpha_3} f_{i_4,\alpha_4}]|_{\text{connected}}\\
        &=\frac{d^4-8d^2+6}{d^2(d^6-14d^4+49d^2-36)}\\
        &[\text{Tr}\left(\rho _{\alpha_1} \rho _{\alpha_2} \rho _{\alpha_3} \rho _{\alpha_4}\right) 
        \times \mathcal{V}_4(\{O_i\})\\
        &+\text{Tr}\left(\{\rho _{\alpha_i} \rho _{\alpha_j} \rho _{\alpha_k} \right\})\times \mathcal{V}_3(\{O_i\}))\\
        &+\mathcal{O}(\frac{1}{d^5}).
    \end{split}
\end{align}

Here, $\mathcal{V}_k$ represents a linear combination of functions that map subsets of $\{O_i\},\{\rho_i\}, i\in[1,\cdots k]$ to scalar values using the $\text{Tr}(\cdot)$ operator.

For large values of $d$, we can approximate by expansion in terms of Hilbert space dimension :
  \begin{equation}
     \mathbb{Q}:=\mathbb{E}[f_1 f_2]=\frac{Q^{[2]}}{d^2}+\frac{Q^{[3]}}{d^3}+\cdots:=\hat{\mathbb{Q}}+\doublehat{\mathbb{Q}}+\mathcal{O}(\frac{1}{d^4})
 \end{equation}
and 
 \begin{equation}
     \mathbb{V}:=\mathbb{E}[f_1 f_2 f_3 f_4]=\frac{V^{[4]}}{d^4}+\frac{V^{[5]}}{d^5}+\cdots:=\hat{\mathbb{V}}+\doublehat{\mathbb{V}}+\mathcal{O}(\frac{1}{d^6})
 \end{equation}
such that for example:
\begin{equation}
    \mathbb{E}[f_{i_1,\alpha_1}f_{i_2,\alpha_2}]|_{\text{conn.}}= \frac{1}{d^2}[\Tr(\rho_{\alpha_1}\rho_{\alpha_2})\Tr(O_{i_1}O_{i_2})]+\mathcal{O}(\frac{1}{d^3})
\end{equation}

To examine the impact of the number of layers, denoted as $L$, we can shift our perspective back to the parametric space. In the classical context, it has been observed that $\mathbb{E}[f_1 f_2]$ can be approximated by averaging over $\mathbb{E}[\langle \nabla_{\boldsymbol{\theta}}f,\nabla_{\boldsymbol{\theta}}f \rangle ]$, based on Bochner's theorem. This concept has been extended to the quantum scenario in \cite{liu2022analytic,liu2022representation}.   For the structure like Eq.\ref{Eq:framework}, if we define:
\begin{equation}
    [\mathbb{H}_{ij,\alpha \beta}]_{\mu \nu}:=\frac{\partial f_{i,\alpha}}{\partial \theta_\mu} \frac{\partial f_{j,\beta}}{\partial \theta_\nu}
\end{equation}
then the evolution of the output can be described by the following differential equation:
\begin{equation}
    df_{i,\alpha}=-\sum_{\mu,\nu} \eta^{\mu \nu} [\mathbb{H}_{ij,\alpha \sigma}]_{\mu\nu}\nabla_{f_{j,\sigma}} \mathcal{L} +\mathcal{O}(\eta^2).
\end{equation}

By defining the left and right operators as $U_{L,\mu}=\prod_{\ell=1}^{\mu-1} W_{\ell}U_{\ell}$, $U_{R,\mu}=\prod_{\ell=\mu+1}^{L}W_{\ell}U_{\ell}$, $V_{L,\mu}=U_{L,\mu}W_{\mu}U_{\mu}$, and $V_{R,\mu}=U_{R,\mu}$, the derivative of the output with respect to a specific parameter (or layer), expressed in a product form, can be represented as follows:
\begin{align}
    \begin{split}
       [\mathbb{H}_{ij,\alpha \beta}]_{\mu\nu} 
       &=\Tr(U_{R,\mu}^\dagger[X_{\mu}, U_{\mu}^\dagger W_{\mu}^\dagger U_{L,\mu}^\dagger O_i U_{L,\mu} W_{\mu}U_\mu]U_{R,\mu}\rho_\alpha)\\
       &\times \Tr(U_{R,\nu}^\dagger[X_{\nu}, U_{\nu}^\dagger W_{\nu}^\dagger U_{L,\nu}^\dagger O_j U_{L,\nu} W_{\nu}U_\nu]U_{R,\nu}\rho_\beta)
    \end{split}
\end{align}

Now, by taking the Haar average over all random unitaries, utilizing Equation \ref{Eq:Haar_average},  We obtain
\begin{align}
    \begin{split}
        &\mathbb{E}[\mathbb{H}_{ij,\alpha \beta}]_{\mu\nu}=\frac{2d}{(d-1)(d+1)(d^2+d)}\times \\
        &[\Tr(O_i O_j)\Tr(\rho_\alpha \rho_\beta)
        -\Tr(O_i)\Tr(O_j)\Tr(\rho_\alpha)\Tr(\rho_\beta)]\\
        &\times \Tr(X_\mu X_\nu)=\frac{c}{d^2} \Tr(O_i O_j)\Tr(\rho_\alpha \rho_\beta) +\mathcal{O}(\frac{1}{d^3})
    \end{split}
\end{align}

where we have defined $c := \frac{\text{Tr}(X_\mu X_\nu)}{d}$, which is typically of order one ($\mathcal{O}(1)$). Based on certain considerations \cite{liu2022analytic}, we can assume that the kernel can be approximated by a frozen one and that the variables $t$ are independent in the later stages of training. For simplicity, let's assume $\eta_{\mu\nu} = \delta_{\mu\nu}$. Then the sum over all parameter sizes $L$ arises as follows:
\begin{equation}
    \sum_{\mu \nu} \eta^{\mu \nu} [H_{ij,\alpha \beta}]_{\mu\nu}\sim \frac{L}{d^2}\Tr(O_i O_j) \Tr(\rho_\alpha \rho_\beta) +\mathcal{O}(\frac{1}{d^3})
\end{equation}

This result motivates us to seek a similar property for the quantum neural network as well, akin to what is observed in classical neural networks:
\begin{align}
    \begin{split}
\mathbb{Q}(\rho_x,\rho_{x'})&=\mathbb{E}_{\theta \sim p(\theta)}[f_\theta(\rho_x) \cdot f_\theta(\rho_{x'})]\\
&\sim \mathbb{E}_{\theta \sim p(\theta)}[\langle \nabla_{\theta}f_{\theta}(\rho_x),\nabla_{\theta}f_{\theta}(\rho_{x'}) \rangle]=\mathbb{Q}_{\text{NTK}}
    \end{split}
\end{align}
Thus, as we will explore in the subsequent section, the impact of depth, denoted as $L$, becomes evident through the faster convergence to a Gaussian distribution:
\begin{eqnarray}
    f(t)-f(0) \propto  (e^{- L \hat{\mathbb{Q}} })^{t\eta}
\end{eqnarray}



\begin{figure}
  \centering
\includegraphics{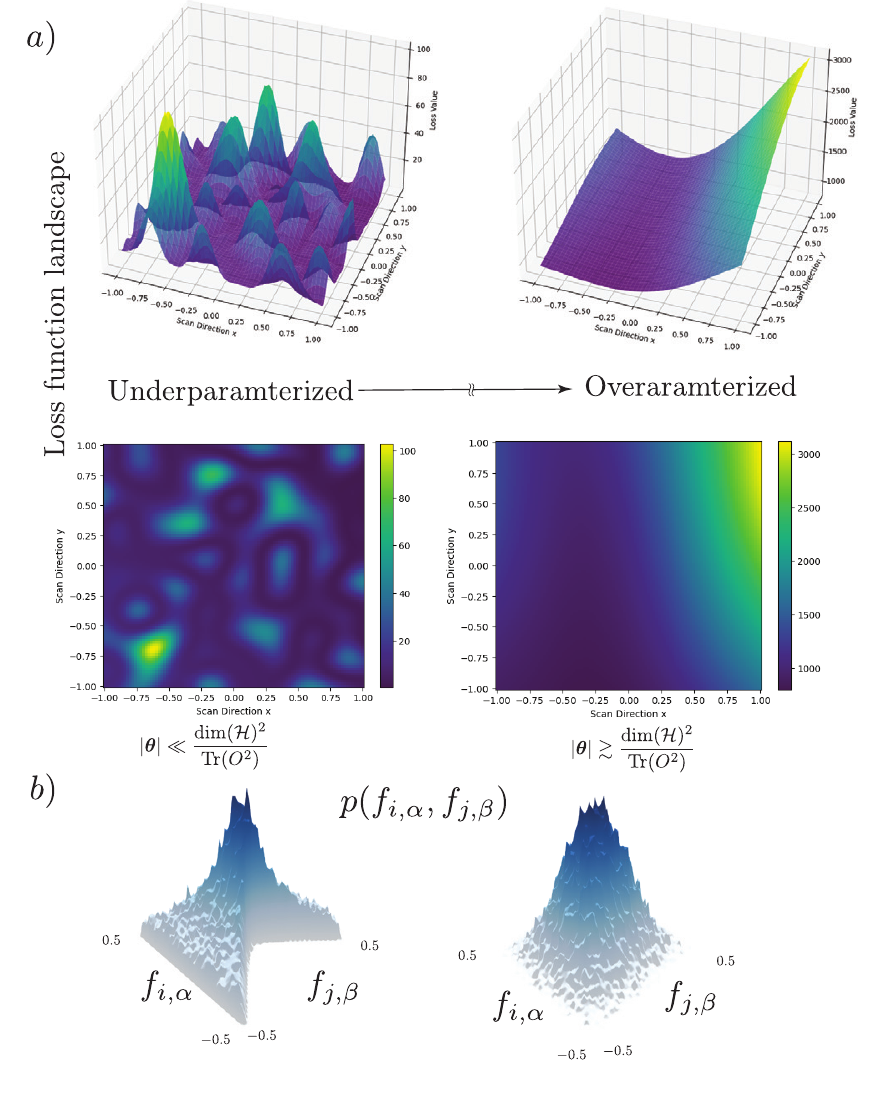}
  \caption{The phase transition of the energy landscape of a quantum neural network (QNN) occurs when transitioning from the underparameterized regime to the overparameterized regime. In the underparameterized regime, where the number of parameters in the QNN is low compared to the dimension of the Hilbert space (or the degrees of freedom), the QNN's highly non-convex loss landscape contains unfavorable local minima with eigenvalues of similar magnitudes as the global minimum. However, as the number of parameters increases, the landscape becomes simpler and exhibits numerous favorable local minima. In this scenario, training and converging towards the actual global minimum become much easier. b) The effect of overparameterization in quantum neural networks (QNNs) is the emergence of Gaussian-like output distributions. As the QNN becomes overparameterized, meaning that the number of parameters exceeds the necessary amount to fit the data, the output distribution of the QNN tends to exhibit Gaussian characteristics. This observation suggests that overparameterization promotes a smoother and more concentrated output distribution, resembling a Gaussian distribution. }
  \label{fig:my_image}
\end{figure}

\section{Bayesian learning or how to get rid of parameters }



For a given quantum model, denoted as $\mathcal{M}$, the circuit architecture associated with it will possess a collection of parameters that may either remain fixed or variable. These variable parameters, referred to as $\boldsymbol{\theta}=\{\theta_i,  i\in[1, \cdots, L]\}$ need to be tuned such that the quantum circuit performs the desired output. 
Since the variational parameters are assumed to be independent of each other, the prior distribution of these parameters is merely the product of the prior distribution of each individual parameter such that $p(\boldsymbol{\theta}|\mathcal{M})=\prod_{\ell=1}^{L}p(\theta_\ell|\mathcal{M})$. 

 Recent investigations have delved into how the prior distribution of the $p(\boldsymbol{\theta}|\mathcal{M})$ as a method of initialization strategies of quantum neural networks, such as Gaussian initialization and reduced-domain parameter initialization \cite{zhang2022escaping}. These studies suggest that such strategies can enhance convergence speed and alleviate the problem of barren plateaus. However, it should be noted that these assumptions do not guarantee that the global minimum can be reached within a reasonable number of iterations using these initialization methods, and there is a risk of getting trapped in undesirable local minima.

\subsection{Marginalization of Parameters }

When considering a specific model $\mathcal{M}$, our prior hypotheses regarding the behavior of the observables we seek to learn are taken into account. The probability distribution describing our prior belief and prediction of the outcome of a quantum circuit, denoted as $p(\mathbb{P}|\boldsymbol{\theta},\mathcal{M})$, can be expressed as follows:

\begin{equation}
    p(f_\mathbb{P}|\mathcal{M})=\int \prod_{i}^L d \theta_{i} \, p(f_\mathbb{P}|\theta_i,\mathcal{M}) \, p(\theta_i|\mathcal{M}) 
\end{equation}

Upon observing the data points of interest $\mathbb{O}$, we can update our beliefs and make predictions for new data points $\mathbb{P}$ using the Bayes' rule:

\begin{equation}
    p(f_{\mathbb{P}}|f_{\mathbb{O}},\mathcal{M})=\int \prod_{i}^L d \theta_{i} \, p(f_{\mathbb{O}}|\theta,\mathcal{M}) \, p(\theta|f_{\mathbb{O}},\mathcal{M})
\end{equation}

However, this equation can be simplified by integrating out the parameters, resulting in the following equation \cite{roberts2022principles}:

\begin{equation}\label{Eq:Bayes_Rule}
    p(f_{\mathbb{P}}|f_{\mathbb{O}},\mathcal{M})=\frac{p(f_{\mathbb{P}},f_{\mathbb{O}}|\mathcal{M})}{p(f_{\mathbb{O}}|\mathcal{M})}
\end{equation}

This formula allows us to update our beliefs and make predictions based on the observed data points without explicitly involving any parameters. In fact, we have bypassed the intermediate step of updating our posterior on the parameters, which can be obtained using different schemes such as Maximum A Posteriori (MAP) or Maximum Likelihood Estimation (MLE). For example, The optimal parameter values with MAP can be found by maximizing the posterior probability
$ \boldsymbol{\theta}^{*}= \underset{\theta}{\arg\max} \, p(\theta| \mathbb{O},\mathcal{M})$.

\section{Probabilistic model of QNN}\label{sec:prob_QNN}

In this section, we present a proposition for a probabilistic approach to quantum neural networks. In this novel viewpoint of quantum neural networks, the output of the circuit is represented as a probability distribution, contrasting the traditional approach where the circuit's output is a delta function across all potential outcomes. The key advantage of this perspective lies in its ability to incorporate uncertainty within the framework, enabling increased flexibility in noise analysis. Furthermore, owing to the inherent probabilistic nature of quantum circuits, the variance from the expectation becomes more pronounced, underscoring its significance.

In general, assuming that the output distribution is symmetric with respect to the origin and neglecting odd-numbered correlation functions, we can describe the output distribution using Equation \ref{p_f}:

In a general framework, assuming symmetry around the origin in the output distribution and excluding correlations of odd order, we can characterize the output distribution as \cite{roberts2022principles}
\begin{equation}\label{p_f}
    p(f)=\frac{e^{-{S}(f)}}{\mathcal{Z}},
\end{equation}
where the action ${S}(\cdot)$ is defined as:
\begin{align}
    \begin{split}
        {S}(f)&=\frac{1}{2}\sum_{\mu,\nu=1}^{n}K^{\mu \nu}f_{\mu}f_{\nu}
        \\
        &+\sum_{m=2}^\Lambda \frac{1}{(2m)!}\sum_{\mu_1, \cdots \mu_{2m}=1}C^{\mu_1 \cdots \mu_{2m}}f_{\mu_1}  \cdots f_{\mu_{2m}}.
    \end{split}
\end{align}
In this formulation, $C^{\mu_1 \cdots \mu_{2m}}$ represents the $m$-th order couplings, and $\Lambda$ serves as the truncation threshold for limiting the series expansion up to the $\Lambda$-th order. Furthermore, the partition function $\mathcal{Z}$ in the denominator of Equation \ref{p_f} is defined as:
\begin{equation}
\mathcal{Z} = \int d^n f  e^{-{S}(f)}
\end{equation}

Here, $C^{\mu_1 \cdots \mu_{2m}}$ represents the $m$-th order couplings defined as
\begin{equation}\label{E_m}
    \mathbb{E}[f_1 f_2 \cdots f_m]=\int df^m p(f) f_1 f_2 \cdots f_m,
\end{equation}
and $\Lambda$ is the limit for truncating the series to the $\Lambda$-th order. The partition function, $\mathcal{Z}$ in the denominator of Equation \ref{p_f}, is given by:
\begin{equation}
    \mathcal{Z}=\int d^n f e^{-\mathcal{S}(f)}
\end{equation}

As the number of qubits, denoted by $n$, increases, the Hilbert space dimension expands exponentially as $d=2^n$. In the asymptotic regime where $d$ approaches infinity, higher-order corrections within the correlation relation become negligible. Specifically, we observe that $\mathbb{E}[f_1 \cdots f_m] \sim \mathcal{O}(1/d^{m'+2})$ for $m'>0$. This implies that the action can be perturbed with respect to the dimension of the Hilbert space:
\begin{equation}
    S = S^{(0)} + \frac{S^{(2)}}{d^2} + \frac{S^{(4)}}{d^4} + \mathcal{O}\left(\frac{1}{d^6}\right).
\end{equation}

This perturbation approach bears resemblance to the $\frac{1}{d}$ expansion technique employed in the domains of statistical mechanics and quantum field theory \cite{roberts2022principles}, where $S^{(0)}$ characterizes the vacuum state of the theory and $S^{(2)}$ represents the first-order interaction and so on.

\subsection{Infinite Width QNN}

In this section, we want to study the extreme over-parameterized case  $ {d\to\infty}$
where all higher order terms than quadratic interaction can be neglected: $S\approx S^{(0)}+S^{(2)}/d^2$. In this regime, it becomes possible to express the output distribution of the quantum neural network  as a Gaussian distribution.

Our learning task, aimed at determining the cachectic of this distribution, entails observing samples from ensembles of data points and subsequently predicting the output of the quantum neural network  for new, unseen data points. We represent the observed datapoints and their corresponding sets as $\mathbb{O}=\{\rho_\beta, f_{i,\beta}:=y_{\beta,i};  \beta \in [1, \cdots |\mathbb{O}|],i\in [1,\cdots, n ]\}$.Here, $|\mathbb{O}|$ denotes the size of the set of observable operators.  In this representation, $\rho_\beta$ denotes the density matrix of the $\beta$-th data point from the observed ensemble, while $y_{i,\beta}:=f_{i,\beta}$ represents the true label of  this particular data point.
Similarly, we define the prediction sets as
$\mathbb{P}=\{\rho_\alpha, =f_{j,\alpha};  \alpha \in [1, \cdots |\mathbb{P}|], j\in [1,\cdots, n]\}$. With these two sets, we can represent our updated posterior as follows:
\begin{equation}\label{exponent}
    p(f_{\mathbb{O}},f_{\mathbb{P}})=\frac{1}{Z} \exp(-\frac{1}{2}\sum_{i,j=1}^n \sum_{\mu_1, \mu_2 \in \mathbb{P}\cup \mathbb{O}} K^{ij,\mu_1 \mu_2}f_{i,\mu_1}f_{j,\mu_2})
\end{equation}
where $Z$ is the normalization factor, given by $|2\pi K|^{n^2}$.

Utilizing Bayes' rule, as presented in Equation \ref{Eq:Bayes_Rule}, we can gain deeper insights into the QNN's output distribution for a particular data point, denoted as $f_{\mathbb{P}}:=f$, based on previously observed data points, $f_{\mathbb{O}}$. Similar computations to those used in classical neural networks, as detailed in \cite{roberts2022principles} and outlined in Appendix \ref{Gaussian_appendix}, allow us to express this distribution as a Gaussian distribution, or potentially as a Gaussian process $p(f)=\frac{1}{Z} \exp(S)$ with Gaussian action:
\begin{align}
    \begin{split}
         p(f)&=\frac{1}{Z} \exp(-S), \\
        S&= \frac{1}{2}\sum_{\substack{i,j=1   \\ \beta_1, \beta_2 \in \mathbb{P}}}^{n}\mathbb{K}^{ij,\beta_1 \beta_2}[f_{i,\beta_1}-m^\infty_{i,\beta_1}][f_{j,\beta_2}-m^\infty_{j,\beta_2}],
    \end{split}
\end{align}
 which is characterized by the following mean and covariance:
\begin{align}
    \begin{split}
         [m]_{i,\mu} &=\sum_{\lambda ,\sigma \in \mathbb{O}}K_{ij,\mu \lambda} K^{ij,\lambda \sigma}y_{i,\sigma}:=K(\mathbb{P},\mathbb{O})K^{-1}(\mathbb{O},\mathbb{O})\boldsymbol{y},\\
        [\mathbb{K}]_{ij,\mu \nu }&= K_{ij,\mu \nu}- \sum_{\lambda, \sigma \in \mathbb{O}}K_{ij,\mu \lambda} K^{ij,\lambda \sigma}K_{ij,\sigma \nu}\\
        &=K(\mathbb{P},\mathbb{O})-K(\mathbb{P},\mathbb{P})K^{-1}(\mathbb{O},\mathbb{O})K(\mathbb{O},\mathbb{P}).
    \end{split}
\end{align}

Here, $K$ is a two-point correlation function representing the kernel function. Therefore, we can establish the explicit relationship between the mean and covariance in terms of the inputs, denoted by 
$\{\rho_{\beta}\}_{\beta\in\mathbb{O}}$ and observables $\{O\}$:

\begin{align}
    \begin{split}
     [m]_{i,\beta}&=\sum_{\alpha_1 ,\alpha_2\in \mathbb{O}}\mathbb{E}[f_{i}f_{j}]_{\beta, \alpha_1}\mathbb{E}[f_{i}f_{j}]^{-1}_{\alpha_1,\alpha_2} y_{i,\alpha_2}\\
    &=\sum_{\alpha_1 ,\alpha_2\in \mathbb{O}} [\Tr(\rho_\beta \rho_{\alpha_1})]
     [\Tr(\rho_{\sharp_1} \rho_{\sharp_2})]^{-1}_{\alpha_1 \alpha_2}y_{i,\alpha_2}
    +\mathcal{O}(\frac{1}{d^2})
        \end{split}
\end{align}

Furthermore, the covariance can be calculated using the following expression:
\begin{align}
    \begin{split}
    &[\mathbb{K}]_{ij,\beta_1 \beta_2 }=
    \frac{\Tr(O_i O_j)}{d^2}[\Tr(\rho_{\beta_1}\rho_{\beta_2})\\
    &-\sum_{\alpha_1, \alpha_2 \in \mathbb{O}}
     \Tr(\rho_{\beta_1}\rho_{\beta_2}) [\Tr(\rho_{\sharp_1}\rho_{\sharp_2})]^{-1}_{\beta_2 \alpha_2} \Tr(\rho_{\alpha_2}\rho_{\beta_2})] +\mathcal{O}(\frac{1}{d^4}).
    \end{split}
\end{align}

In literature, the term $\Tr(\rho_x \rho_x)$ is occasionally denoted as the Quantum fidelity kernel $K^{\mathbb{F}}(x,x')$ \cite{huang2021power}. Using this notation, we can make the previous equation more understandable: 
\begin{equation}
    \mathbb{K}_{ij,\mathbb{P}}=\frac{\Tr(O_i O_j)}{d^2} [K^{\mathbb{F}}_{\mathbb{P},\mathbb{O}}- K^\mathbb{F}_{\mathbb{P},\mathbb{P}}(K^\mathbb{F})^{-1}_{_{\mathbb{O},\mathbb{O}}}K^\mathbb{F}_{\mathbb{O},\mathbb{P}}] +\mathcal{O}(\frac{1}{d^4})
\end{equation}

\subsection{Parameter space representation}

Within the context of parameter space, our learning endeavor can be seen as the creation of a parametric model $f_{\text{QNN}}(\rho; \boldsymbol{\theta})$, accompanied by a distribution $p(\boldsymbol{\theta})$. The aim is to locate the best value of $\boldsymbol{\theta}=\boldsymbol{\theta}^{*}$, ensuring minimal error considering available computational resources (like the number of qubits, gates, and observed datapoints). The goal is to have our model closely approximate the actual distribution and align accurately with the real datapoints $f_{\text{QNN}}(x;\boldsymbol{\theta}^{*})\approx f(x)$. The aspiration is that, with proper initialization near the actual global minimum, we can locate this global minimum using gradient descent methods. As quantum neural networks  lean towards overparameterization, this assumption becomes more credible. Given this assumption, the output of the Quantum Neural Network (QNN) can be approximated using a Taylor series:
\begin{align}\label{Eq:Taylor_expansion}
    \begin{split}
       & f^{i,(t)}_{\alpha}= f^{i,(0)}_{\alpha}
       -\eta \sum_{ij,\beta} \mathbb{H}^{ij,\beta \alpha}\epsilon_{j,\beta}\leftrightarrow \mathbb{Q}_\text{NTK}\\
        &+\frac{\eta^2}{2}\sum_{i j_1 j_2, \beta_1 \beta_2}d\mathbb{H}^{ij_1 j_2,\alpha \beta_1 \beta_2}\epsilon_{j_1,\beta_1}\epsilon_{j_2,\beta_2}\leftrightarrow \text{d}\mathbb{Q}_\text{NTK}\\
        &-\frac{\eta^3}{6}\sum_{i j_1 \cdots \alpha_3} dd\mathbb{H}^{i j_1 \cdots \alpha_3}\epsilon_{j_1 \beta_1}\epsilon_{j_2 \beta_2}\epsilon_{j_3 \beta_3}\leftrightarrow \text{dd}\mathbb{Q}_\text{NTK}\\
        &+\mathcal{O}(\frac{1}{d^2})
    \end{split}
\end{align}
where $\epsilon= \nabla_f \mathcal{L}$ denotes the derivative of the loss function with respect to the output function and $\mathbb{H}$ is defined as:
\begin{equation}
    \mathbb{H}_{ij,\alpha \beta}=\sum_{\mu \nu}\eta_{\mu\nu}\frac{\partial f_{i,\alpha}}{\partial \theta_\mu}\frac{\partial f_{j,\beta}}{\partial \theta_\nu}.
\end{equation}

As the width of the Quantum Neural Network (QNN) approaches infinity, $\mathbb{H}$ can be approximated by a frozen kernel, referred to as the Quantum Neural Tangent Kernel (QNTK):
\begin{equation}
     \mathbb{H}_{ij,\alpha \beta}=\mathbb{Q}_{ij,\alpha \beta}+\mathcal{O}({\frac{1}{d^2}})
\end{equation}

By comparing the expansion of the parameter space with the functional method outlined in Sec.\ref{sec:prob_QNN}, we can draw the following analogy:
\begin{align}
    \begin{split}
         \mathbb{Q}_{\text{NTK}}&: \mathbb{E}[\frac{\partial f_{i_1, \alpha_1}}{\partial \theta_\mu}\frac{\partial f_{j,\alpha_2}}{\partial \theta_\nu}]
         \sim\frac{1}{d^2}\Tr(\rho_{\alpha_1} \rho_{\alpha_2})\Tr(O_iO_j) \\&
         \sim \mathbb{E}[f_{i_1, \alpha_1}f_{j,\alpha_2}]={K}_{ij,\alpha_1 \alpha_2 }
    \end{split}
\end{align}
Likewise, for the other terms in the Eq.\ref{Eq:Taylor_expansion} expansion, these are known as quantum meta kernels, which correspond to the third and fourth-order correlations:
\begin{align}
    \begin{split}
        \text{d}\mathbb{Q}_{\text{NTK}}&: \mathbb{E}[\frac{\partial^2 f_{i_1,\alpha_1}}{\partial \theta_\mu \partial \theta_\nu}\frac{\partial f_{i_2,\alpha_2}}{\partial \theta_\lambda}\frac{\partial f_{i_3,\alpha_3}}{\partial \theta_\sigma}]=0\\
        &
        \sim \mathbb{E}[f_{i_1,\alpha_1}f_{i_2,\alpha_2}f_{i_3,\alpha_3}],
    \end{split}
\end{align}
and
\begin{align}
    \begin{split}
        \text{dd}\mathbb{Q}_\text{NTK}&: \mathbb{E}[\frac{\partial^3 f}{\partial \theta_1 \partial \theta_2 \partial \theta_3}\frac{\partial f}{\partial \theta_4}\frac{\partial f}{\partial \theta_5}\frac{\partial f}{\partial \theta_6}]\\
      & + \mathbb{E}[\frac{\partial^2 f}{\partial \theta_1 \partial \theta_2 }\frac{\partial^2 f}{\partial \theta_3 \partial \theta_4 }\frac{\partial f}{\partial \theta_5}\frac{\partial f}{\partial \theta_6}]\\
       &\leftrightarrow \mathbb{E}[f_{i_1,\alpha_1}f_{i_2,\alpha_2}f_{i_3,\alpha_3}f_{i_4,\alpha_4}]=\mathbb{V}_{i_1 i_2 i_3 i_4, \alpha_1 \alpha_2 \alpha_3 \alpha_4}
    \end{split}
\end{align}

Solving Eq.\ref{Eq:Taylor_expansion} entails dealing with a complex set of differential equations, which generally do not yield a closed-form solution \cite{roberts2022principles}. However, in the limit of $d \rightarrow \infty$, it is acceptable to only retain the linear term. The solution to the differential equation, in terms of steps (represented here by the continuous variable $t$), can be expressed as:

\begin{align}
    \begin{split}
    f_{i,\alpha}^{(t)}& \simeq m^{(t)}_{i,\alpha}+\Sigma^{(t)}_{i,\alpha}
    =\sum_{\beta_1, \beta_2 \in \mathbb{O}}\mathbb{Q}_{\alpha \beta_1} \mathbb{Q}^{\beta_1 \beta_2}(1-e^{-\eta \mathbb{Q}t})y_{i,\beta_2}\\
    &+f_{i,\alpha}^{(0)}-\sum_{\beta_1, \beta_2 \in \mathbb{O}}\mathbb{Q}_{\alpha \beta_1} \mathbb{Q}^{\beta_1 \beta_2}(1-e^{-\eta \mathbb{Q}t}) f_{i,\beta_2}^{(0)}.
    \end{split}
\end{align}

In the limit of large iteration times, we can determine the output of the circuit (without actually performing any training) as a linear kernel regression:
\begin{equation}
    f_{\text{QNN}}(\rho_x) \approx f_{\mathbb{Q}_{\text{NTK}}}=\mathbb{Q}_{\text{NTK}}(\rho_x,\rho_X)^T \cdot \mathbb{Q}_{\text{NTK}}(\rho_X,\rho_X)^{-1}\cdot Y,
\end{equation}

and the optimized parameters at the  end of training  would be determined as 
\begin{equation}
\boldsymbol{\theta}^{*}=\boldsymbol{\theta}^{(0)}-\eta \nabla_{\boldsymbol{\theta}}f\cdot \mathbb{Q}_{\text{NTK}}^{-1}(\boldsymbol{f}-\boldsymbol{y})
\end{equation}

To see the connection with the Gaussian process, let's consider the expected output for a particular input $\rho_\alpha$, which can be given as:
\begin{equation}
    \mathbb{E}[f_{i,\alpha}^{(t)}]=\sum_{\beta_1, \beta_2 \in \mathbb{O}}\mathbb{Q}_{\alpha \beta_1}\mathbb{Q}^{\beta_1 \beta_2}(1-e^{-\eta \mathbb{Q}t})y_{i,\beta_2},
\end{equation}
which, in the limit of infinite iterations, converges to $m_{i,\alpha}^{(\infty)}:=\sum_{\beta_1, \beta_2 \in \mathbb{O}}\mathbb{Q}_{\alpha \beta_1}\mathbb{Q}^{\beta_1 \beta_2}y_{i,\beta_2}$.

Likewise, the covariance of the circuit's output will take the following form:
\begin{align}
    \begin{split}
       & \mathbb{E}[f_{i,\alpha_1}^{(t)} f_{j,\alpha_2}^{(t)}]|_{\text{connected}}=\mathbb{Q}_{\alpha_1 \alpha_2}
       \\
       &-\sum_{\beta_1,\beta_2 \in \mathbb{P}}\mathbb{Q}_{\alpha_2 \beta_1} \mathbb{Q}^{\beta_1 \beta_2}(1-e^{-\eta \mathbb{Q}t})\mathbb{Q}_{\alpha_1\beta_2}\\
        &-\sum_{\beta_1,\beta_2 \in \mathbb{P}}\mathbb{Q}_{\alpha_1 \beta_1} \mathbb{Q}^{\beta_1 \beta_2}(1-e^{-\eta \mathbb{Q}t})\mathbb{Q}_{\alpha_2\beta_2}
        \\&+\sum_{\beta_1,\cdots, \beta_4 \in \mathbb{P}}\mathbb{Q}_{\alpha_1 \beta_1}\mathbb{Q}^{\beta_1 \beta_2 }[1-e^{-\eta \mathbb{Q}t}]\mathbb{Q}_{\alpha_2 \beta_3}[1-e^{-\eta \mathbb{Q}t}]\\
        &\times \mathbb{Q}^{\beta_3 \beta_4}\mathbb{Q}_{\beta_4 \beta_2}
        :=\Sigma_{\alpha_1 \alpha_2}^{ij,t}.
    \end{split}
\end{align}

The higher order moments are of the order $\mathcal{O}(1/d^2)$ and can be disregarded. Based on this observation, we can infer that the most likely distribution to describe the statistics of the output is a Gaussian process. Upon comparing with the functional description of the output distribution, we observe that we obtain the same distribution as predicted by the Gaussian process, particularly as $t \rightarrow\infty$, and substituting $\mathbb{Q}$ with $\mathbb{K}$:
\begin{equation}
    f^{(t)}\sim \mathcal{GP}(m^{(t)},\Sigma^{(t)}) \stackrel{\lim\limits_{t \rightarrow \infty} }{\rightarrow} 
\mathcal{GP}(m^{(\infty)},\mathbb{K})
\end{equation}

comparing the linear kernel regression method with the Gaussian process,

\begin{align}
    \begin{split}
        f \sim f_{\mathbb{Q}_{\text{NTK}}} &\leftrightarrow  \text{Kernel Regression}\\
        p(f) \sim \mathcal{N}(\mu,\Sigma)&\leftrightarrow \text{Gaussain Process}
    \end{split}
\end{align}

\subsection{Finite Width QNN}

When working within the finite width regime of the parameter space, it becomes necessary to consider higher-order terms, such as dQNTK and ddQNTK, which are based on the higher-order derivative terms dH and dHH. Using traditional methods to obtain higher-order corrections in the output distribution, such as for dQNTK, can be unfeasible and not provide a closed-form solution. However, in this section, we demonstrate that by utilizing Bayesian learning, we can obtain higher-order corrections of the output distribution in a closed form, at least for the case of dQNTK.

To enhance the model's ability to capture the finite width regime, it becomes necessary to go beyond the Gaussian distribution and utilize representation learning techniques. By incorporating additional correlation functions into the model, it becomes better equipped to accurately represent the underlying reality.

To obtain a nearly-Gaussian (NG) distribution, comparable to ddQNTK, we need to incorporate the next leading order in finite width, which is of $O(1/d)$. This is because in the nearly-Gaussian distribution, the odd terms are zero or equivalently, dQNTK is zero.

\begin{equation}
{S}_{\text{NG}}:=\frac{1}{2}\sum_{\mu,\nu}^{n}K^{\mu \nu}f_{\mu}f_{\nu}-\frac{\lambda}{4!}\sum_{\nu_1 \nu_2 \nu_3 \nu_4=1}^{n} V^{\nu_1 \nu_2 \nu_3 \nu_4} f_{\nu_1}f_{\nu_2}f_{\nu_3}f_{\nu_4}.
\end{equation}
such that
\begin{align}
    \begin{split}
        K^{\alpha_1 \alpha_2}&=\hat{\mathbb{Q}}^{\alpha_1 \alpha_2}+\mathcal{O}(\frac{1}{d^3})\\
        V^{\alpha_1 \alpha_2\alpha_3 \alpha_4}&=\hat{\mathbb{V}}^{(\alpha_1 \alpha_2)(\alpha_3 \alpha_4)}+\mathcal{O}(\frac{1}{d^5})
    \end{split}
\end{align}

By employing the same calculations as we did for the infinite width scenario, we can determine the contributions of finite width structures to the mean and covariance:

\begin{align}
    \begin{split}
        &\mathbb{E}[f_{i,\beta}]
        =m^{\infty}_{i,\beta}+\frac{1}{3!}\sum_{j_1,\beta \in \mathbb{P}} K_{ij_1,\beta \beta_1}\\
    &\times[\sum_{\substack{ j_2 j_3 j_4 \\
        \alpha_2 \alpha_3 \alpha_4 \in \mathbb{O}}}V^{j_1 j_2 j_3 j_4, \beta \alpha_2 \alpha_3 \alpha_4} y_{j_2, \alpha_2} y_{j_3, \alpha_3} y_{j_4, \alpha_4}]\\
        &\sum_{\substack{ j_1 j_2 j_3 \\
        \beta_1 \beta_2 \beta_3 \in \mathbb{P}}} [K_{i j_1, \beta \beta_1}K_{j_2 j_3, \beta_2 \beta_3}+K_{i j_2, \beta \beta_2}K_{j_1 j_3, \beta_1 \beta_3}
        \\
        &+K_{i j_3, \beta \beta_3}K_{j_1 j_2, \beta_1 \beta_2}] \sum_{j_4, \alpha \in \mathbb{O} }V^{j_1 j_2 j_3 j_4, \beta_1 \beta_2 \beta_3 \alpha} y_{j_4,\alpha}\\
        &+\mathcal{O}(\frac{1}{d^4}),
    \end{split}
\end{align}
and
 \begin{align}
    \begin{split}
      &\Sigma'= \mathbb{E}[f_{i_1,\mu_1} f_{i_2,\mu_2} ]|_{\text{conn.}}=K_{i_1 i_2,\mu_1 \mu_2}
       \\
       &-\frac{1}{2}
     \sum_{\substack{j_1 j_2 j_3 j_4=1   \\ \nu_1 \nu_2 \nu_3 \nu_4}}
      V^{j_1 j_2 j_3 j_4,\nu_1 \nu_2 \nu_3 \nu_4} K_{i_1 j_1,\mu_1 \nu_1}K_{i_2 j_2,\mu_2 \nu_2}\\
      &\times K_{j_3 j_4,\nu_3 \nu_4}
    \end{split},
\end{align}
 where $m^{(\infty)}_{i,\alpha}=\sum_{\beta_1, \beta_2}K_{\alpha_1 \beta_1}K^{\beta_1 \beta_2}y_{i,\beta_2 }$ thus e can refine the kernel regression method by incorporating finite-size corrections: 
 \begin{equation}
     m'=\bar{f}= m^{(\infty)}+ m^{(1/d)}+ \mathcal{O}(\frac{1}{d^2})
 \end{equation}

If we aim to maintain the Gaussian approximation, the corrected covariance, which arises from higher orders, and the effects of wiring with others, can be described as 
\begin{equation}
    f\approx \mathcal{GP}(m',\Sigma').
\end{equation}

However, and in the large $d$ limit and the output distribution can be approximate better by forth order correlator as  
:
\begin{equation}
    p(f)=p(f| y_\alpha, \hat{\mathbb{Q}}_{\mu \nu}, \hat{\mathbb{V}}_{\mu\nu \delta \sigma }).
\end{equation}
such that $\hat{\mathbb{Q}}=\Sigma'$ and $\hat{\mathbb{V}}$ given by (see appendix for more details)
\begin{align}
    \begin{split}
        &\mathbb{V}_{i_1 i_2 i_3 i_4, \mu_1 \mu_2 \mu_3 \mu_4 }=\sum_{j_1 j_2 j_3 j_4 }\sum_{\nu_1\nu_2 \nu_3 \nu_4} V^{j_1 j_2 j_3 j_4,\nu_1\nu_2 \nu_3 \nu_4}\\
        &\times K_{i_1 j_1,\mu_1 \nu_1}K_{i_2 j_2,\mu_2 \nu_2}K_{i_3 j_3, \mu_3 \nu_3 }K_{i_4 j_4,\mu_4 \nu_4}+\mathcal{O}(\frac{1}{d^5})
    \end{split}
\end{align}

\section{Conclusion}
In this study, we studied the  variational quantum circuits, which serve as a model for quantum neural networks from the lense of function space, in stead of parmatr space. We explored the impact of over-parameterization through the expansion of width and depth . We obsereved theortically and by simulation that this led to a shift in the output distribution of the quantum models towards a Gaussian process. We also examined how the finite size of the width can skew the distribution towards non-Gaussianity. We demonstrated how this can be achieved using a $1/d$ expansion in closed form relation, setting the stage for systematic consideration of higher-order corrections.

Note added:  This work was presented informally to the scientific community during the APS March Meeting 2023, and toward the completion of this work, we became aware of work with a similar topic on arXiv \cite{garcia2023deep}.

\bibliography{main}
\bibliographystyle{plain}

\begin{widetext}
\appendix


\section{Infinite width QNN}\label{Gaussian_appendix}

In the context of infinite-width quantum neural networks, the prior distribution takes the form of a simple Gaussian distribution. Consequently, the posterior distribution can be represented as follows:

\begin{equation}\label{exponent}
    p(f_{\mathbb{O}},f_{\mathbb{P}})=\frac{1}{\sqrt{|2\pi K|^{n^2}}} \exp(-\frac{1}{2}\sum_{i,j=1}^n \sum_{\mu_1, \mu_2 \in \mathbb{P}\cup \mathbb{O}} K^{ij,\mu_1 \mu_2}f_{i,\mu_1}f_{j,\mu_2})
\end{equation}

 It is possible to divide the series in the exponent into four separate matrices. Starting with 
 \begin{equation}
      K_{ij,\mu_1 \mu_2}=\begin{bmatrix} 
	K_{ij,o_1 o_2} & K_{ij,o_1 p_2}  \\
	K_{ij,p_1 o_2} & K_{ij,p_1 p_2} 
 \end{bmatrix}=
 \begin{bmatrix} 
	\mathbb{E}[f_{i,o_1}f_{j,o_2}]& \mathbb{E}[f_{i,o_1}f_{j,p_2}]  \\
	\mathbb{E}[f_{i,p_1}f_{j,o_2}] & \mathbb{E}[f_{i,p_1}f_{j,p_2}],
 \end{bmatrix}
 \end{equation}

The inverse of $K_{\mu_1 \mu_2}$, for a specific $ij$ index, is expressed as $\sum_{\mu_2 } K^{\mu_1 \mu_2} K_{\mu_2 \mu_3}=\delta_{\mu_1,\mu_3}$. This separation for the first-order correction is valid since $K_{ij,\alpha \beta}\sim \frac{1}{d^2}\Tr(O_i O_j)\Tr(\rho_\alpha \rho_\beta):=M_{ij} G_{\alpha \beta}$ This expression can be broken down into four distinct sectors as follows:
\begin{equation}
   K^{\mu_1 \mu_2}= \begin{bmatrix}
       K^{\alpha_1 \alpha_2} & K^{\alpha_1 \beta_2}\\
       K^{\beta_1 \alpha_2} & K^{\beta_1 \beta_2}
   \end{bmatrix}=
   \begin{bmatrix}
        K^{\alpha_1 \alpha_2}+\sum_{\alpha_3 \alpha_4 \in \mathbb{O}, \beta_3,\beta_4 \in \mathbb{P}} K^{\alpha_1 \alpha_3} K_{\alpha_3 \beta_3} \mathbb{K}^{\beta_3 \beta_4} K_{\beta_4 \alpha_4} K^{\alpha_4 \alpha_2}  & -\sum_{\alpha_3 \in \mathbb{O}, \beta_3 \in \mathbb{P}}{K}^{\alpha_1 \alpha_3} K_{\alpha_3 \beta_3} \mathbb{K}^{\beta_3 \beta_2}\\
        -\sum_{\alpha_3 \in \mathbb{O}, \beta_3 \in \mathbb{P}}\mathbb{K}^{\beta_1 \beta_3} K_{\beta_3 \alpha_3} K^{\alpha_3 \alpha_2} & \mathbb{K}^{\beta_1 \beta_2},
    \end{bmatrix}
\end{equation}
where in the above notation, $\mathbb{K}^{p p}$ is defined as $(K_{pp}-K_{po}K_{oo}^{-1}K_{op})^{-1}$, and the expression for $\mathbb{K}{p_1 p_2}$ is $K{p_1 p_2}- \sum_{o_3, o_4 \in \mathbb{O}}K_{p_1 o_3} K^{o_3 o_4}K_{o_4 p_2}$.


Now, We can rewrite the exponent in Eq.\ref{exponent} using the $K^{\mu_1 \mu_2}$ matrix:
\begin{align}\label{quad}
    \begin{split}
        &-\frac{1}{2}\sum_{i,j=1}^n \sum_{\mu_1, \mu_2 \in \mathbb{P}\cup \mathbb{O}} K^{ij,\mu_1 \mu_2}f_{i,\mu_1}f_{j,\mu_2}=-\frac{1}{2}\sum_{ij=1}^n \sum_{\beta_1, \beta_2 \in \mathbb{P}}\mathbb{K}^{ij,\beta_1 \beta_2}f_{i,\beta_1}f_{j,\beta_2}-\frac{1}{2}\sum_{i,j=1}^n \sum_{\alpha_1, \alpha_2 \in \mathbb{O}} K^{ij,\alpha_1 \alpha_2}y_{i,\alpha_1}y_{j,\alpha_2}\\
        &-\frac{1}{2}\sum_{i,j=1}^n \sum_{\alpha_1, \in \mathbb{O}, \beta_1 \in \mathbb{P}} K^{ij,\alpha_1 \beta_1} f_{i,\beta_1} y_{j,\alpha_1}-\frac{1}{2}\sum_{i,j=1}^n \sum_{\alpha_1, \in \mathbb{O}, \beta_1 \in \mathbb{P}} K^{ij,\beta_1 \alpha_1} f_{j,\beta_1} y_{i,\alpha_1}\\
        &=-\frac{1}{2}\sum_{i,j=1}^{n}\sum_{\beta_1,\beta_2 \in \mathbb{P}} \mathbb{K}^{ij,\beta_1 \beta_2}[f_{i,\beta_1}-\sum_{\alpha_3, \alpha_4 \in \mathbb{O}}K_{ij,\beta_1 \alpha_3}K^{ij,\alpha_3 \alpha_4}y_{j,\alpha_4}][f_{i,\beta_2}-\sum_{\alpha'_3, \alpha'_4 \in \mathbb{O}}K_{ij,\beta_2 \alpha'_3}K^{ij,\alpha'_3 \alpha'_4}y_{j,\alpha'_4}]\\
        &-\frac{1}{2}\sum_{ij=1}^n \sum_{\alpha_1, \alpha_2 \in \mathbb{O}} K^{ij,\alpha_1 \alpha_2}y_{i,\alpha_1}y_{j,\alpha_2}-\frac{1}{2}\sum_{i,j=1}^n \sum_{\beta_1, \beta_2 \in \mathbb{O}}\mathbb{K}^{ij,\beta_1 \beta_2}(\sum_{\alpha_3, \alpha_4 \in \mathbb{O}}K_{ij,\beta_1 \alpha_3}K^{ij,\alpha_3 \alpha_4}y_{i,\alpha_4})\\
        &\times(\sum_{\alpha'_3, \alpha'_4 \in \mathbb{O}}K_{ij,\beta_2 \alpha'_3}K^{ij,\alpha'_3 \alpha'_4}y_{j,\alpha'_4}).
    \end{split}
\end{align}

By defining $m_{i,\beta}^{\infty}=\sum_{\alpha_1 \alpha_2}K_{ij,\beta \alpha_1} K^{ij,\alpha_1 \alpha_2}y_{j,\alpha_2}$, we can ultimately express the posterior distribution Eq.\ref{exponent} as:
\begin{align}
    \begin{split}
    p(f):=p(f_{\mathbb{O}},f_{\mathbb{P}})&=\frac{1}{\sqrt{|2\pi K|^{n}}} \exp(-\frac{1}{2}\sum_{ij=1}^n \sum_{\alpha_1, \alpha_2 \in \mathbb{O}} K^{ij,\alpha_1 \alpha_2}y_{i,\alpha_1}y_{j,\alpha_2}-\frac{1}{2}\sum_{ij=1}^n \sum_{\beta_1, \beta_2 \in \mathbb{O}}\mathbb{K}^{ij,\beta_1 \beta_2}m^\infty_{i,\beta_1}m^\infty_{j,\beta_2})\\
    &\times -\frac{1}{2}\sum_{ij=1}^{n}\sum_{\beta_1,\beta_2 \in \mathbb{P}} \mathbb{K}^{ij,\beta_1 \beta_2}[f_{i,\beta_1}-m^\infty_{i,\beta_1}][f_{j,\beta_2}-m^\infty_{j,\beta_2}]
    \end{split}
\end{align}

The aforementioned joint posterior distribution lets us ascertain our desired posterior distribution for unobserved data $f \in \mathbb{O}$, as outlined in Eq.\ref{Eq:Bayes_Rule}:
\begin{equation}
    p(f| f_{\mathbb{O}})=C_{(2)}(f_{\mathbb{O}})\times \exp[-\frac{1}{2}\sum_{ij=1}^{n}\sum_{\beta_1,\beta_2 \in \mathbb{P}} \mathbb{K}^{ij,\beta_1 \beta_2}[f_{i,\beta_1}-m^\infty_{i,\beta_1}][f_{j,\beta_2}-m^\infty_{j,\beta_2}]],
\end{equation}
where $C(f_{\mathbb{O}})$ is solely a function of the previously observed data, $f_{\mathbb{O}}$, and can be considered as a constant.

Therefore, we can deduce that in the case of an infinitely deep scenario, the output of the QNN follows a Gaussian process with the respective mean and variance:
\begin{align}
    \begin{split}
         p(f)& \sim \mathcal{GP}(m,\mathbb{K}),\\
         [m]_{i,\mu} &=\sum_{\lambda ,\sigma \in \mathbb{O}}K_{ij,\mu \lambda} K^{ij,\lambda \sigma}y_{i,\sigma},\\
        [\mathbb{K}]_{ij,\mu \nu }&= K_{ij,\mu \nu}- \sum_{ij,\lambda, \sigma \in \mathbb{O}}K_{ij,\mu \lambda} K^{ij,\lambda \sigma}K_{ij,\sigma \nu}.
    \end{split}
\end{align}

or equivalently 
\begin{align}
\begin{split}
        [m]_{i,\beta} &=\sum_{\alpha_1 ,\alpha_2 \in \mathbb{O}}K_{ij,\beta \alpha_1} K^{ij,\alpha_1 \alpha_2}y_{i,\alpha_2}=\sum_{\alpha_1 ,\alpha_2\in \mathbb{O}}\mathbb{E}[f_{i}f_{j}]_{\beta, \alpha_1}\mathbb{E}[f_{i}f_{j}]^{-1}_{\alpha_1,\alpha_2} y_{i,\alpha_2}\\
        &=\sum_{\alpha_1 ,\alpha_2\in \mathbb{O}} [\frac{1}{d^2}\Tr(O_iO_j)\Tr(\rho_\beta \rho_{\alpha_1})][\frac{1}{d^2}\Tr(O_iO_j)\Tr(\rho_{\sharp_1} \rho_{\sharp_2})]^{-1}_{\alpha_1 \alpha_2}y_{i,\alpha_2}+\mathcal{O}(\frac{1}{d^2})
        \\
        [\mathbb{K}]_{ij,\beta_1 \beta_2 }&= K_{ij,\beta_1 \beta_2}- \sum_{\alpha_1, \alpha_2 \in \mathbb{O}}K_{ij,\beta_1 \beta_2} K^{ij,\beta_2 \alpha_2}K_{ij,\alpha_2 \beta_2}\\
        &=\mathbb{E}[f_{i}f_{j}]_{\beta_1\beta_2}-\sum_{\alpha_1, \alpha_2 \in \mathbb{O}}\mathbb{E}[f_{i}f_{j}]_{\beta_1\beta_2}\mathbb{E}[f_{i}f_{j}]^{-1}_{\beta_2\alpha_2}
        \mathbb{E}[f_{i}f_{j}]_{\alpha_2 \beta_2}\\
        &=\frac{\Tr(O_i O_j)}{d^2}[\Tr(\rho_{\beta_1}\rho_{\beta_2})-\sum_{\alpha_1, \alpha_2 \in \mathbb{O}} \Tr(\rho_{\beta_1}\rho_{\beta_2}) \Tr(\rho_{\sharp_1}\rho_{\sharp_2})]^{-1}_{\beta_2 \alpha_2} \Tr(\rho_{\alpha_2}\rho_{\beta_2})] +\mathcal{O}(\frac{1}{d^4})\\
        &:=\frac{\Tr(O_i O_j)}{d^2} [K^\mathbb{Q}_{\beta_1 \beta_2}- \sum_{\alpha_1, \alpha_2 \in \mathbb{O}}K^\mathbb{Q}_{\beta_1 \beta_2} (K^\mathbb{Q})^{-1}_{\beta_12\alpha_2} K^\mathbb{Q}_{\alpha_2 \beta_2}] +\mathcal{O}(\frac{1}{d^4})
\end{split}
\end{align}

\section{Beyond Gaussian Distribution and $\frac{1}{d}$ Expansion}

Here, we utilize the notation as prescribed by \cite{roberts2022principles} and reproduce the calculation performed in that textbook for our quantum scenario, where the k-point functions may not exhibit perfect symmetry. 

This is due to the fact that despite the orthonormal random matrix which can be written as sum over pairings, this theorem doesn't hold to the random unitary in general

This comes from the observation that 

\begin{align}
    \begin{split}
        \mathbb{E}[f_1 \cdots f_k]&=\mathbb{E}[\Tr(U_1 \rho_1 U^\dagger O_1)\cdots \Tr(U_k \rho_k U^\dagger O_k)]
    \end{split}
\end{align}

[NEED TO BE COMPLETE]

By assuming the parity symmetry of the output , $f$, with respect to zero, we can describe the output distribution by series of the even moments(correlation function) denominated as $p(f)$. These moments are defined as:
\begin{equation}\label{E_M}
    \mathbb{E}[f_1 f_2 \cdots f_{k}]=\int df^k p({f}) f_1 f_2 \cdots f_{k}
\end{equation}
which is zero for while is$k$ is an odd number. Based on these $k-$point correlators, the distributions can be defined as follows:
\begin{equation}\label{p_f}
    p({f})=\frac{e^{-S(f)}}{\mathcal{Z}}
\end{equation}
such that the action, ${S}(\cdot)$ is given by:
\begin{equation}
    {S}({f})=\frac{1}{2}\sum_{\mu,\nu=1}^{n}K^{\mu \nu}f_{\mu}f_{\nu}+\sum_{m=2}^\Lambda \frac{1}{(2m)!}\sum_{\mu_1, \cdots \mu_{2m}=1}{C}^{\mu_1 \cdots \mu_{2m}}f_{\mu_1}  \cdots f_{\mu_{2m}},
\end{equation}
Here, $  C^{\mu_1 \cdots \mu_{2m}}$ is the $m$-th order couplings, and $\Lambda$ is the practical limit for cutting the series and limiting it to some $\Lambda$-th order. The partition function, $\mathcal{Z}$ in the denominator in Eq.\ref{p_f}is defined as:
\begin{equation}
    \mathcal{Z}=\int d^n f e^{-{S}({f})}
\end{equation}. 

An action with a small fourth-order interaction can provide the first leading order of deviation from Gaussian distribution. This nearly Gaussian action can be defined as follows:
\begin{equation}
{S}_{\text{NG}}:=\frac{1}{2}\sum_{\mu,\nu=1}^{n}K^{\mu \nu}f_{\mu}f_{\nu}-\frac{\lambda}{4!}\sum_{\nu_1 \nu_2 \nu_3 \nu_4=1}^{n} V^{\nu_1 \nu_2 \nu_3 \nu_4} f_{\nu_1}f_{\nu_2}f_{\nu_3}f_{\nu_4} +\mathcal{O}({\lambda^2}).
\end{equation}

We can take this one step further and construct actions beyond nearly Gaussian, and so forth:
\begin{align}
    \begin{split}
        {S}_{\text{NNG}}:=&\frac{1}{2}\sum_{\mu,\nu=1}^{n}K^{\mu \nu}f_{\mu}f_{\nu}-\frac{\lambda}{4!}\sum_{\nu_1 \nu_2 \nu_3 \nu_4=1}^{n} V^{\nu_1 \nu_2 \nu_3 \nu_4} f_{\nu_1}f_{\nu_2}f_{\nu_3}f_{\nu_4} \\
        &+\frac{\lambda^2}{4!}\sum_{\nu_1 \nu_2 \nu_3 \nu_4,\nu_5,\nu_6=1}^{n} U^{\nu_1 \nu_2 \nu_3 \nu_4 \nu_5 \nu_6} f_{\nu_1}f_{\nu_2}f_{\nu_3}f_{\nu_4}f_{\nu_5}f_{\nu_6}+\mathcal{O}({\lambda^3})
    \end{split}
\end{align}

The partition function of the nearly Gaussian action can be calculated by:
\begin{align}
    \begin{split}
        \mathcal{Z}_{\text{NG}}&=\int \prod_{\mu}df_{\mu}e^{-{S}({f})}=\sqrt{|2\pi K|}\mathbb{E}_K[e^{-\frac{\lambda}{4!}\sum_{\nu_1 \nu_2 \nu_3 \nu_4=1}^{n} V^{\nu_1 \nu_2 \nu_3 \nu_4} f_{\nu_1}f_{\nu_2}f_{\nu_3}f_{\nu_4}}  ] \\
        &= \sqrt{|2\pi K|}\mathbb{E}_K[ 1-\frac{\lambda}{4!}\sum_{\nu_1 \nu_2 \nu_3 \nu_4=1}^{n} V^{\nu_1 \nu_2 \nu_3 \nu_4} f_{\nu_1}f_{\nu_2}f_{\nu_3}f_{\nu_4}   +\mathcal{O}(\lambda^2) ]\\
        &=\sqrt{|2\pi K|}[1-\frac{3\lambda}{4!}\sum_{\nu_1 \nu_2 \nu_3 \nu_4}V^{\nu_1 \nu_2 \nu_3 \nu_4} K_{\nu_1 \nu_2 }K_{\nu_3\nu_4} +\mathcal{O}(\lambda^2)]
    \end{split}
\end{align}
In the third line, we used the Wicks theorem which holds for Gaussian distribution:
\begin{equation}
    \mathbb{E}_K[f_1 \cdots f_{2k}]=\sum_{\text{all pairs}} \mathbb{E}_{K}[f_{i_1}f_{j_1}]\cdots  \mathbb{E}_{K}[f_{i_k}f_{j_k}]
\end{equation}



the second term in the nearly-Gaussian action can be re-written in terms of $\Delta$:

\begin{align}
    \begin{split}
        S_2&=\frac{1}{4!}[\sum_{\substack{
        j_1 j_2 j_3 j_4\\
        \mu_1 \mu_2 \mu_3 \mu_4  \mathbb{O}\cup \mathbb{P}}}V^{j_1 j_2 j_3 j_4, \mu_1 \mu_2 \mu_3 \mu_4}f_{j_1, \mu_1}f_{j_2, \mu_2}f_{j_3, \mu_3}f_{j_4, \mu_4}]\\
        &=\frac{1}{4!}[\sum_{\substack{
        j_1 j_2 j_3 j_4\\
        \mu_1 \mu_2 \mu_3 \mu_4  \mathbb{O}}}V^{j_1 j_2 j_3 j_4, \mu_1 \mu_2 \mu_3 \mu_4}y_{j_1, \mu_1}y_{j_2, \mu_2}y_{j_3, \mu_3}y_{j_4, \mu_4}]\\
        &+\sum_{j_1 j_2 j_3 j_4 }\sum_{\beta_1 \beta_2 \beta_3 \beta_4 \in \mathbb{P}} V^{j_1 j_2 j_3 j_4, \beta_1 \beta_2 \beta_3 \beta_4}f_{j_1,\beta_1}f_{j_2,\beta_2}f_{j_3,\beta_3}f_{j_4,\beta_4} \\
        &+4\sum_{\substack{j_1, j_2 \\
        \beta_1 \beta_2 \in \mathbb{P}}}f_{j_1,\beta_1}f_{j_2,\beta_2}
        \sum_{\substack{j_3 j_4 \\
       \alpha_3 \alpha_4 \in \mathbb{O} }}
        V^{j_1 j_2 j_3 j_4, \beta_1 \beta_2 \alpha_3 \alpha_4}y_{j_3, \alpha_3}y_{j_4, \alpha_4}\\
        &4 \sum_{\substack{ j_1 j_2 j_3 \\
        \beta_1 \beta_2 \beta_3 \in \mathbb{P}}} f_{j_1, \beta_1} f_{j_2, \beta_2} f_{j_3, \beta_3} \sum_{j_4, \alpha \in \mathbb{O} }V^{j_1 j_2 j_3 j_4, \beta_1 \beta_2 \beta_3 \alpha} y_{j_4,\alpha}\\
        &4 \sum_{j_1, \beta \in \mathbb{P}} f_{j_1, \beta} \sum_{\substack{ j_2 j_3 j_4 \\
        \alpha_2 \alpha_3 \alpha_4 \in \mathbb{O}}}V^{j_1 j_2 j_3 j_4, \beta \alpha_2 \alpha_3 \alpha_4} y_{j_2, \alpha_2} y_{j_3, \alpha_3} y_{j_4, \alpha_4}
    \end{split}
\end{align}

Now, with the explicit relations for $S_1$ and $S_2$, we can derive the correlation functions in terms of $K$, $V$, and the $y$ variables.
similar to the classical one, by adhering to the notation and methodology in \cite{roberts2022principles}, we can write the first-order correlation function, the mean, as:
\begin{align}
    \begin{split}
        \mathbb{E}[f_{i,\beta}]&=\int df p(f|y)f_{i,\beta}=\int df \frac{e^{-S_1}}{Z}[e^{-S_2} f_{i,\beta}]=\int df\frac{e^{-S_1}}{Z'}[e^{-S_2(f-m^{\infty})} ((f_{i,\beta}-m^{\infty}_{i,\beta})+m^{\infty}_{i,\beta})]\\
        &=m^{\infty}_{i,\beta}+\mathbb{E}_{S_1}[(f_{i,\beta}-m^{\infty}_{i,\beta})e^{S_2(f-m^{\infty})}]\\
        &=m^{\infty}_{i,\beta}+\frac{1}{3!}\sum_{j_1,\beta \in \mathbb{P}} K_{ij_1,\beta \beta_1}[\sum_{\substack{ j_2 j_3 j_4 \\
        \alpha_2 \alpha_3 \alpha_4 \in \mathbb{O}}}V^{j_1 j_2 j_3 j_4, \beta \alpha_2 \alpha_3 \alpha_4} y_{j_2, \alpha_2} y_{j_3, \alpha_3} y_{j_4, \alpha_4}]\\
        &\frac{1}{3!}\sum_{\substack{ j_1 j_2 j_3 \\
        \beta_1 \beta_2 \beta_3 \in \mathbb{P}}} [K_{i j_1, \beta \beta_1}K_{j_2 j_3, \beta_2 \beta_3}+K_{i j_2, \beta \beta_2}K_{j_1 j_3, \beta_1 \beta_3}+K_{i j_3, \beta \beta_3}K_{j_1 j_2, \beta_1 \beta_2}] \sum_{j_4, \alpha \in \mathbb{O} }V^{j_1 j_2 j_3 j_4, \beta_1 \beta_2 \beta_3 \alpha} y_{j_4,\alpha}
    \end{split}
\end{align}

we can evaluate the second moment using the following equation:
\begin{align}\label{Eq:two_point}
    \begin{split}
        \mathbb{E}[f_{\mu_1} f_{\mu_2} ]&=\frac{1}{\mathcal{Z}}\int \prod_{\mu}df_{\mu}e^{-{S}({f})} f_{\mu_1} f_{\mu_2}=\frac{\sqrt{|2\pi K|}}{\mathcal{Z}}\mathbb{E}_K[ f_{\mu_1} f_{\mu_2} e^{-\frac{ \lambda}{4!}\sum_{\nu_1 \nu_2 \nu_3 \nu_4}V^{\nu_1 \nu_2 \nu_3 \nu_4} f_{\nu_1}f_{\nu_2}f_{\nu_3}f_{\nu_4}}]\\
        &=\frac{1}{\mathcal{Z}}[\mathbb{E}_K[ f_{\mu_1}f_{\mu_2}]-\frac{\lambda}{4!}\sum_{\nu_1 \nu_2 \nu_3 \nu_4}V^{\nu_1 \nu_2 \nu_3 \nu_4} \mathbb{E}_K[f_{\mu_1}f_{\mu_2}f_{\nu_1}f_{\nu_2}f_{\nu_3}f_{\nu_4} ] +\mathcal{O}(\lambda^2)]\\
         &=K_{\mu_1 \mu_2}-\frac{\lambda}{2}\sum_{\nu_1 \nu_2 \nu_3 \nu_4} V^{\nu_1 \nu_2 \nu_3 \nu_4} K_{\mu_1 \nu_1}K_{\mu_2 \nu_2}K_{\nu_3 \nu_4}+\mathcal{O}(\lambda^2),
    \end{split}
\end{align}
and we obtain a similar outcome for the connected two-point correlator because:
\begin{eqnarray}
     \mathbb{E}[f_{\mu_1} f_{\mu_2} ]|_{\text{connected}}= \mathbb{E}[f_{\mu_1} f_{\mu_2} ]-\mathbb{E}[f_{\mu_1}]\mathbb{E}[f_{\mu_2}]=\mathbb{E}[f_{\mu_1} f_{\mu_2} ].
\end{eqnarray}
Now, by incorporating the measurement indices, we get:
\begin{equation}
     \mathbb{E}[f_{i_1,\mu_1} f_{i_2,\mu_2} ]|_{\text{conn.}}=K_{i_1 i_2,\mu_1 \mu_2}-\frac{\lambda}{2}
     \sum_{\substack{j_1 j_2 j_3 j_4=1   \\ \nu_1 \nu_2 \nu_3 \nu_4}}
      V^{j_1 j_2 j_3 j_4,\nu_1 \nu_2 \nu_3 \nu_4} K_{i_1 j_1,\mu_1 \nu_1}K_{i_2 j_2,\mu_2 \nu_2}K_{j_3 j_4,\nu_3 \nu_4}+\mathcal{O}(\lambda^2)
\end{equation}

Our subsequent action entails computing the fourth moment:
 \begin{align}
     \begin{split}
         \mathbb{E}[f_{\mu_1}f_{\mu_2}f_{\mu_3}f_{\mu_4}]&=\int \prod_{\mu}df_{\mu}e^{-\mathcal{S}({f})} f_{\mu_1} f_{\mu_2}f_{\mu_3} f_{\mu_4}\\
         &=\frac{\sqrt{|2\pi K|}}{\mathcal{Z}}[\langle f_{\mu_1}f_{\mu_2}f_{\mu_3}f_{\mu_4}\rangle_K- \frac{\lambda}{4!}\sum_{\nu_1 \nu_2 \nu_3 \nu_4} V^{\nu_1 \nu_2 \nu_3 \nu_4}\langle f_{\mu_1}f_{\mu_2}f_{\nu_1}f_{\nu_2}f_{\nu_3}f_{\nu_4}\rangle_K +\mathcal{O}(\lambda^2)]\\
         &=\frac{\sqrt{|2\pi K|}}{\mathcal{Z}}[\big(\sum_{\text{all 3 pairing}}K_{\mu_{\ell_1}\mu_{\ell_2}}K_{\mu_{\ell_3}\mu_{\ell_4}})-\frac{\lambda}{4!}\sum_{\text{all 15 pairing}}K_{\mu_{\ell_1}\mu_{\ell_2}}\cdots K_{\mu_{\ell_{5}}\mu_{\ell_6}}+\mathcal{O}(\lambda^2)]\\
         &=[1-\frac{3\lambda}{4!}\sum_{\nu_1 \nu_2 \nu_3 \nu_4}V^{\nu_1 \nu_2 \nu_3 \nu_4} K_{\nu_1 \nu_2 }K_{\nu_3\nu_4} +\mathcal{O}(\lambda^2)]\times[\big(\sum_{\text{all 3 pairing}}K_{\mu_{\ell_1}\mu_{\ell_2}}K_{\mu_{\ell_3}\mu_{\ell_4}})\\
         &-\frac{\lambda}{4!}\sum_{\text{all 15 pairing}}K_{\mu_{\ell_1}\mu_{\ell_2}}\cdots K_{\mu_{\ell_{5}}\mu_{\ell_6}}+\mathcal{O}(\lambda^2)]\\
         &=K_{\mu_1 \mu_2}K_{\mu_3 \mu_4} -\frac{\lambda}{4!}\sum_{\nu_1\nu_2\nu_3 \nu_4}V^{\nu_1 \nu_2 \nu_3 \nu_4}[12 K_{\mu_1 \nu_1}K_{\mu_2 \nu_2}K_{\mu_3 \mu_4}K_{\nu_3 \nu_4}+12 K_{\mu_3 \nu_1}K_{\mu_4 \mu_2}K_{\mu_1 \mu_2}K_{\nu_3\nu_4}]\\
         &+K_{\mu_1 \mu_3}K_{\mu_2 \mu_4} -\frac{\lambda}{4!}\sum_{\nu_1\nu_2\nu_3 \nu_4}V^{\nu_1 \nu_2 \nu_3 \nu_4}[12 K_{\mu_1 \nu_1}K_{\mu_3 \nu_3}K_{\mu_2 \mu_4}K_{\nu_3 \nu_4}+12 K_{\mu_2 \nu_1}K_{\mu_4 \mu_3}K_{\mu_1 \mu_3}K_{\nu_3\nu_4}]\\
         &+K_{\mu_1 \mu_4}K_{\mu_2 \mu_3} -\frac{\lambda}{4!}\sum_{\nu_1\nu_2\nu_3 \nu_4}V^{\nu_1 \nu_2 \nu_3 \nu_4}[12 K_{\mu_1 \nu_1}K_{\mu_4 \nu_3}K_{\mu_2 \mu_3}K_{\nu_3 \nu_4}+12 K_{\mu_2 \nu_1}K_{\mu_3 \mu_4}K_{\mu_1 \mu_4}K_{\nu_3\nu_4}]\\
         &-\lambda \sum_{\nu_1\nu_2 \nu_3 \nu_4} V^{\nu_1\nu_2 \nu_3 \nu_4}K_{\mu_1 \nu_1}K_{\mu_2 \nu_2}K_{\mu_3 \nu_3 }K_{\mu_4 \nu_4}+\mathcal{O}(\lambda^2)\\
         &=\mathbb{E}[f_{\mu_1}f_{\mu_2}]\mathbb{E}[f_{\mu_3}f_{\mu_4}]+\mathbb{E}[f_{\mu_1}f_{\mu_3}]\mathbb{E}[f_{\mu_2}f_{\mu_4}]+\mathbb{E}[f_{\mu_1}f_{\mu_4}]\mathbb{E}[f_{\mu_2}f_{\mu_3}]\\
         &-\lambda \sum_{\nu_1\nu_2 \nu_3 \nu_4} V^{\nu_1\nu_2 \nu_3 \nu_4}K_{\mu_1 \nu_1}K_{\mu_2 \nu_2}K_{\mu_3 \nu_3 }K_{\mu_4 \nu_4}+\mathcal{O}(\lambda^2),
     \end{split}
 \end{align}
Then, after incorporating the measurement indices, we obtain:
\begin{equation}\label{Eq:four_point}
     \mathbb{E}[f_{i_1,\mu_1}f_{i_2,\mu_2}f_{i_3,\mu_3}f_{i_4,\mu_4}]|_{\text{connected}}=-\lambda \sum_{\substack{j_1 j_2 j_3 j_4=1   \\ \nu_1 \nu_2 \nu_3 \nu_4}} V^{j_1 j_2 j_3 j_4,\nu_1\nu_2 \nu_3 \nu_4}K_{i_1 j_1,\mu_1 \nu_1}K_{i_2 j_2,\mu_2 \nu_2}K_{j_3 j_4,\mu_3 \nu_3 }K_{\mu_4 \nu_4}+\mathcal{O}(\lambda^2)
 \end{equation}

With the relation in Eq.\ref{Eq:two_point} and Eq.\ref{Eq:four_point} we can Find explicit form of $K$ and $V$:
\begin{equation}
    \mathbb{Q}_{\mu_1 \mu_2}^{i_1 i_2}=K_{\mu_1 \mu_2}^{i_1 i_2}-\frac{\lambda}{2}\sum_{j_1 j_2 j_3 j_4}\sum_{\nu_1 \nu_2 \nu_3 \nu_4} V^{\nu_1 \nu_2 \nu_3 \nu_4}_{j_1 j_2 j_3 j_4} K_{\mu_1 \nu_1}^{j_1 i_1}K_{\mu_2 \nu_1}^{j_2 i_2}K_{\nu_3 \nu_4}^{j_3 i_3 }+\mathcal{O}(\lambda^2)
\end{equation}
 and 
 \begin{equation}
     \mathbb{V}_{\mu_1 \mu_2 \mu_3 \mu_4 }=-\lambda \sum_{\nu_1\nu_2 \nu_3 \nu_4} V^{\nu_1\nu_2 \nu_3 \nu_4}K_{\mu_1 \nu_1}K_{\mu_2 \nu_2}K_{\mu_3 \nu_3 }K_{\mu_4 \nu_4}+\mathcal{O}(\lambda^2)
 \end{equation}

 \begin{equation}
     \mathbb{V}_{i_1 i_2 i_3 i_4, \mu_1 \mu_2 \mu_3 \mu_4 }=-\lambda \sum_{j_1 j_2 j_3 j_4 }\sum_{\nu_1\nu_2 \nu_3 \nu_4} V^{j_1 j_2 j_3 j_4,\nu_1\nu_2 \nu_3 \nu_4}K_{i_1 j_1,\mu_1 \nu_1}K_{i_2 j_2,\mu_2 \nu_2}K_{i_3 j_3, \mu_3 \nu_3 }K_{i_4 j_4,\mu_4 \nu_4}+\mathcal{O}(\lambda^2)
 \end{equation}

\section{Haar average  }

In this section, we will demonstrate how to compute the two-point correlation functions as a quantum kernel.
\begin{eqnarray}
    \mathbb{Q}:=\mathbb{E}_{p\sim \text{Haar}(U(\boldsymbol{\theta}))}[f_{i_1,\alpha_1}(U(\boldsymbol{\theta}))f_{i_2,\alpha_2}(U(\boldsymbol{\theta}))]
\end{eqnarray}
and four point correlation function as  quantum kernel
Quantum meta Kernel
\begin{eqnarray}
    \mathbb{V}:=\mathbb{E}_{p\sim \text{Haar}(U(\boldsymbol{\theta}))}[f_{i_1,\alpha_1}(U(\boldsymbol{\theta}))f_{i_2,\alpha_2}(U(\boldsymbol{\theta})
    )
    f_{i_3,\alpha_3}(U(\boldsymbol{\theta}))f_{i_4,\alpha_4}(U(\boldsymbol{\theta})
    )]
\end{eqnarray}

and how get express them  as series of Hilbert space's dimension, $d$:

  \begin{equation}
      \mathbb{Q}_{i_1 i_2,\mu_1 \mu_2}=\mathbb{E}[f_{i_1,\mu_1}f_{i_2,\mu_2}]=\frac{Q^{[2]}_{i_1 i_2,\mu_1 \mu_2}}{d^2}+\frac{Q^{[3]}_{i_1 i_2,\mu_1 \mu_2}}{d^3}+\cdots:=\hat{\mathbb{Q}}_{i_1 i_2,\mu_1 \mu_2}+\doublehat{\mathbb{Q}}_{i_1 i_2,\mu_1 \mu_2}+\mathcal{O}(\frac{1}{d^4})
 \end{equation}

 and Similarly

 \begin{equation}
    \mathbb{V}_{i_1 i_2 i_3 i_4, \mu_1 \mu_2 \mu_3 \mu_4}=\mathbb{E}[f_{i_1,\mu_1} f_{i_2,\mu_2} f_{i_3,\mu_3} f_{i_4,\mu_4}]=\frac{V^{[4]}_{\sharp}}{d^4}+\frac{V^{[5]}_{\sharp}}{d^5}+\cdots:=\hat{\mathbb{V}}_{\sharp}+\doublehat{\mathbb{V}}_{\sharp}+\mathcal{O}(\frac{1}{d^6})
 \end{equation}

\subsection{Two point correlation}\label{section:sub_Haar_two}

In order to calculate the two-point correlation, for $t-$design circuit, as discusse the main text, it is necessary to perform Haar averaging over terms such as $\Tr(\cdot)\Tr(\cdot)$. To transform this form into the identity described in Equation \ref{Eq:Haar_average}, we can define projection operators $P_{r_1 r_2}$ in the following manner:

\begin{align}
    \begin{split}
        \Tr(\rho_1 U O_1 U^\dagger )\Tr(\rho_2 U O_2 U^\dagger )&=\sum_{i_1}\sum_{i_2} \langle i_1 | \rho_1 U O_1 U^\dagger | i_1\rangle \langle i_2 | \rho_2 U O_2 U^\dagger |i_2 \rangle=\sum_{i_1, i_2}\Tr(P_{21}\rho_1 U O_1 U^\dagger P_{12}\rho_2 U O_2 U^\dagger)\\
    &:=\sum_{i_1, i_2}\Tr(\tilde{\rho_1} U O_1 U^\dagger \tilde{\rho_2} U O_2 U^\dagger),
    \end{split}
\end{align}
where we have defined $P_{12}:=|i_1\rangle \langle i_2 |$, $P_{21}:=|i_2\rangle \langle i_1 |$ and $P \rho:=\tilde{\rho}$. Then according to Eq.\ref{Eq:Haar_average}
\begin{align}\label{Eq:D2}
    \begin{split}
         \mathbb{E}[\Tr(\tilde{\rho_1} U O_1 U^\dagger \tilde{\rho_2} U O_2 U^\dagger)]&=\frac{1}{d^2-1}[\Tr(O_1)\Tr(O_2)\Tr(\tilde{\rho_1}\tilde{\rho_2})+\Tr(O_1 O_2)\Tr(\tilde{\rho_1})\Tr(\tilde{\rho_2})]\\
         &-\frac{1}{d^3-d}[\Tr(O_1 O_2)\Tr(\tilde{\rho_1}\tilde{\rho_2})+\Tr(O_1)\Tr(O_2)\Tr(\tilde{\rho_1})\Tr(\tilde{\rho_2}))]
    \end{split}
\end{align}

Now, based on the given definition, we can rephrase $\tilde{\rho}$ as $\rho$ as a following:
\begin{equation}
    \Tr(\tilde{\rho_1}\tilde{\rho_2})=\Tr(P_{21}\rho_1 P_{12}\rho_2 )=\langle i_1 |\rho_1 |i_1 \rangle \langle i_2 |\rho_2 |i_2 \rangle \rightarrow \sum_{i_1,i_2}\Tr(\tilde{\rho_1}\tilde{\rho_2})=\Tr(\rho_1)\Tr(\rho_2).
\end{equation}
Similarly, we can apply the same rephrasing to other term:
\begin{equation}
\Tr(\tilde{\rho_1})\Tr(\Tilde{\rho_2})=\Tr(P_{21}\rho_1 )\Tr(P_{12}\rho_2 )=\langle i_1 |\rho_1 |i_2 \rangle \langle i_2 |\rho_2 |i_1 \rangle \rightarrow \sum_{i_1 i_2}\Tr(\tilde{\rho_1})\Tr(\Tilde{\rho_2}) =\Tr(\rho_1 \rho_2 )
\end{equation}

Therefore, Equation \ref{Eq:D2} can be expressed using the actual density matrix as follows:

\begin{align}
    \begin{split}
    \mathbb{E}[\Tr(\rho_1 U O_1 U^\dagger )\Tr(\rho_2 U O_2 U^\dagger )]&=
        \sum_{i_1, i_2 }\mathbb{E}[\Tr(\tilde{\rho_1} U O_1 U^\dagger \tilde{\rho_2} U O_2 U^\dagger)]\\
        &=\frac{1}{d^2-1}[\Tr(O_1)\Tr(O_2)
         \Tr(\rho_1)\Tr(\rho_2)+\Tr(O_1 O_2)\Tr(\rho_1 \rho_2)]\\
         &-\frac{1}{d^3-d}[\Tr(O_1 O_2)\Tr(\rho_1)\Tr(\rho_2)+\Tr(O_1)\Tr(O_2)\Tr(\rho_1 \rho_2))].
    \end{split}
\end{align}
To calculate the connected form of the two-point correlator, we can utilize the fact that the expectation value of odd correlations is zero. Similar to the previous identity, we obtain the following expression:
\begin{align}
    \begin{split}
        \mathbb{E}[f_{i_1,\alpha_1}f_{i_2,\alpha_2}]|_{\text{connected}}&=\mathbb{E}[f_{i_1,\alpha_1}f_{i_2,\alpha_2}]-\mathbb{E}[f_{i_1,\alpha_1}]\mathbb{E}[f_{i_2,\alpha_2}]\\
    &=\frac{\text{Tr}\left(O_{i_2}O_{i_1}\right)}{d-d^3}+\frac{\text{Tr}\left(O_{i_2}O_{i_1}\right)\text{Tr}\left(\rho_{\alpha_1}\rho_{\alpha_2}\right)}{d^2-1}+\frac{\text{Tr}\left(O_{i_1}\right)\text{Tr}\left(O_{i_2}\right)}{d^2-1}\\
    &+\frac{\text{Tr}\left(O_{i_1}\right)\text{Tr}\left(O_{i_2}\right) \text{Tr}\left(\rho _{\alpha_1}\right) \text{Tr}\left(\rho_{\alpha_2}\right)}{d^2}.
    \end{split}
\end{align}
At large $d$, the two point correlator exhibits the following behavior:
\begin{equation}
    \mathbb{E}[f_{i_1,\alpha_1}f_{i_2,\alpha_2}]|_{\text{connected}}= \frac{1}{d^2}[\Tr(\rho_{\alpha_1}\rho_{\alpha_2})\Tr(O_{i_1}O_{i_2})]+\mathcal{O}(\frac{1}{d^3})
\end{equation}

\subsection{Four-point correlator }
We can employ the same method discussed in Section \ref{section:sub_Haar_two} to derive closed-form expressions for the four-point correlation function. We begin by considering the four-point interaction term:
\begin{align}
    \begin{split}
        &\Tr(\rho_1 U O_1 U^\dagger )\Tr(\rho_2 U O_2 U^\dagger )
        \Tr(\rho_3 U O_3 U^\dagger )
        \Tr(\rho_4 U O_4 U^\dagger )\\
        &=\sum_{i_1,i_2,i_3, i_4}\Tr(P_{41}\rho_1 U O_1 U^\dagger P_{12}\rho_2 U O_2 U^\dagger P_{23}\rho_3 U O_3 U^\dagger P_{34}\rho_3 U O_4 U^\dagger)\\
        &=\sum_{i_1,i_2,i_3, i_4} \Tr(\tilde{\rho_1}U O_1 U^\dagger  \tilde{\rho_2}U O_2 U^\dagger \tilde{\rho_3}U O_3 U^\dagger 
        \tilde{\rho_4}U O_4 U^\dagger ).
    \end{split}
\end{align}
such that the  projection operators maps the density matrix to
\begin{equation}
\tilde{\rho}_1=P_{41}\rho_1=|i_4\rangle \langle i_1| \rho_1\,\,\,\, \tilde{\rho}_2=P_{12}\rho_2=|i_1\rangle \langle i_2| \rho_2,\,\,\,\, ,\tilde{\rho}_3=P_{23}\rho_3=|i_2\rangle \langle i_3| \rho_3\,\,\,
,\tilde{\rho}_4=P_{34}\rho_4=|i_3\rangle \langle i_4| \rho_4.
\end{equation}

By this definition, we can observe that, for example, $\Tr(\tilde{\rho}_1 \tilde{\rho}_2\tilde{\rho}_3 \tilde{\rho}_4)=\Tr(\rho_1)\Tr(\rho_2)\Tr(\rho_3)\Tr(\rho_4)=1$. Similarly, $\Tr(\tilde{\rho}_1\tilde{\rho}_4\tilde{\rho}_3\tilde{\rho}_2)=\Tr(\rho_1 \rho_3)\Tr(\rho_2 \rho_4)$, and so on. Now by employing Eq.\ref{Eq:Haar_average} we get the bare four point correlator as follows with the help of  \cite{fukuda2019rtni} package:

\begin{align}
    \begin{split}
        &\int_{\mathcal{N}(d)} \Tr(U^\dagger O U XU^\dagger O U Y U^\dagger O U Z U^\dagger O UW)d\mu(U)=\frac{d^4-8d^2+6}{d^2(d^6-14d^4+49d^2-36)}\times [\\
        &\text{Tr}\left(O_1\right) \text{Tr}\left(O_2\right) \text{Tr}\left(O_3\right) \text{Tr}\left(O_4\right) \text{Tr}(XYZW)+\text{Tr}\left(O_2O_1\right) \text{Tr}\left(O_3\right) \text{Tr}\left(O_4\right) \text{Tr}(Y) \text{Tr}(XZW)\\
    &\text{Tr}\left(O_3O_2\right) \text{Tr}\left(O_1\right) \text{Tr}\left(O_4\right) \text{Tr}(Z) \text{Tr}(XYW)+\text{Tr}\left(O_3 O_1\right) \text{Tr}\left(O_2\right)\text{Tr}\left(O_4\right) \text{Tr}(XW) \text{Tr}(ZY)\\
        &\text{Tr}\left(O_2O_3O_1\right)\text{Tr}\left(O_4\right) \text{Tr}(Y) \text{Tr}(Z) \text{Tr}(XW)+\text{Tr}\left(O_3O_2O_1\right) \text{Tr}\left(O_4\right) \text{Tr}(XZYW)\\
    &\text{Tr}\left(O_4O_2\right) \text{Tr}\left(O_1\right)\text{Tr}\left(O_3\right) \text{Tr}(WZ) \text{Tr}(XY)+\text{Tr}\left(O_4 O_1\right) \text{Tr}\left(O_2\right) \text{Tr}\left(O_3\right)\text{Tr}(X) \text{Tr}(ZWY)\\
&\text{Tr}\left(O_2O_4O_1\right) \text{Tr}\left(O_3\right) \text{Tr}(X) \text{Tr}(Y) \text{Tr}(WZ)+\text{Tr}\left(O_4O_2O_1\right) \text{Tr}\left(O_3\right)\text{Tr}(XZWY)\\
&\text{Tr}\left(O_4O_3O_1\right) \text{Tr}\left(O_2\right) \text{Tr}(XWYZ)+\text{Tr}\left(O_4O_3\right) \text{Tr}\left(O_1\right) \text{Tr}\left(O_2\right)\text{Tr}(W) \text{Tr}(XYZ)\\
&\text{Tr}\left(O_4O_3O_2\right) \text{Tr}\left(O_1\right) \text{Tr}(XYWZ)+\text{Tr}\left(O_3O_4O_1\right) \text{Tr}\left(O_2\right) \text{Tr}(W) \text{Tr}(X)\text{Tr}(ZY)\\
&\text{Tr}\left(O_4 O_3 O_2 O_1\right) \text{Tr}(WY) \text{Tr}(XZ)+\text{Tr}\left(O_3O_4O_2\right) \text{Tr}\left(O_1\right) \text{Tr}(W) \text{Tr}(Z) \text{Tr}(XY)\\
&\text{Tr}\left(O_3 O_4 O_2 O_1\right) \text{Tr}(W) \text{Tr}(XZY)+\text{Tr}\left(O_4 O_2O_3 O_1\right) \text{Tr}(Z) \text{Tr}(XWY)\\
&\text{Tr}\left(O_2 O_4 O_3 O_1\right) \text{Tr}(Y) \text{Tr}(X W Z)+\text{Tr}\left(O_3 O_2 O_4 O_1\right) \text{Tr}(X) \text{Tr}(W Z Y)\\
&\text{Tr}\left(O_3O_1\right) \text{Tr}\left(O_4 O_2\right) \text{Tr}(X W Z Y)+\text{Tr}\left(O_2 O_3 O_4 O_1\right) \text{Tr}(W) \text{Tr}(X) \text{Tr}(Y) \text{Tr}(Z)\\
    &\text{Tr}\left(O_3O_2\right) \text{Tr}\left(O_4O_1\right) \text{Tr}(X) \text{Tr}(Z) \text{Tr}(W Y)+\text{Tr}\left(O_2 O_1\right) \text{Tr}\left(O_4 O_3\right) \text{Tr}(W) \text{Tr}(Y) \text{Tr}(XZ)] \\
        &+\frac{1}{d^5-10 d^3+9 d}\times [\cdots \text{ 49 terms }\cdots ]\\
        &+\frac{3 \left(2 d^2-3\right)}{d^2 \left(d^6-14 d^4+49 d^2-36\right)}\times[\cdots \text{ 52 terms }\cdots ]\\
        &+\frac{1}{d^7-14 d^5+49 d^3-36 d}\times[\cdots \text{ 53 terms }\cdots].
    \end{split}
\end{align}

Then, according to the Wick theorem: 
\begin{equation}
    \mathbb{E}[f_1 f_2 f_3 f_4]|_{\text{connected}}=\mathbb{E}[f_1 f_2 f_3 f_4]-\mathbb{E}[f_1 f_2]\mathbb{E}[f_3 f_4]-\mathbb{E}[f_1 f_3]\mathbb{E}[f_2 f_4]-\mathbb{E}[f_1 f_4]\mathbb{E}[f_2 f_3]
\end{equation}
Hence, the connected four-point correlator can be expressed as: 
\begin{align}
    \begin{split}
        &\mathbb{E}[\Tr(\rho_{\alpha_1} U O_i U^\dagger)\Tr(\rho_{\alpha_2} U O_j U^\dagger)\Tr(\rho_{\alpha_3} U O_k U^\dagger)\Tr(\rho_{\alpha_4} U O_\ell U^\dagger)]|_{\text{connected}}\\
        &=\mathbb{E}[\Tr(\rho_{\alpha_1} U O_i U^\dagger)\Tr(\rho_{\alpha_2} U O_j U^\dagger)\Tr(\rho_{\alpha_3} U O_k U^\dagger)\Tr(\rho_{\alpha_4} U O_\ell U^\dagger)]\\
        &-\mathbb{E}[\Tr(\rho_{\alpha_1} U O_i U^\dagger) \Tr(\rho_{\alpha_2} U O_j U^\dagger)]\mathbb{E}[\Tr(\rho_{\alpha_3} U O_k U^\dagger) \Tr(\rho_{\alpha_4} U O_\ell U^\dagger)]\\
        &-\mathbb{E}[\Tr(\rho_{\alpha_1} U O_i U^\dagger) \Tr(\rho_{\alpha_3} U O_k U^\dagger)]\mathbb{E}[\Tr(\rho_{\alpha_2} U O_j U^\dagger) \Tr(\rho_{\alpha_4} U O_\ell U^\dagger)]\\
        &-\mathbb{E}[\Tr(\rho_{\alpha_1} U O_i U^\dagger) \Tr(\rho_{\alpha_4} U O_\ell U^\dagger)]\mathbb{E}[\Tr(\rho_{\alpha_2} U O_j U^\dagger) \Tr(\rho_{\alpha_3} U O_k U^\dagger)].
    \end{split}
\end{align}
This leads to a closed-form relation for the connected four-point correlator:
\begin{align}
    \begin{split}
        \mathbb{E}[f_{i_1,\alpha_1} f_{i_2,\alpha_2} f_{i_3,\alpha_3} f_{i_4,\alpha_4}]|_{\text{connected}}&=\frac{d^4-8d^2+6}{d^2(d^6-14d^4+49d^2-36)}\times[\\
        & \text{Tr}\left(\rho_{\alpha_1} \rho_{\alpha_2} \rho _{\alpha_3} \rho_{\alpha_4}\right)\times[\text{Tr}\left(O_{i_1} O_{i_2} O_{i_3} O_{i_4}\right)+\text{Tr}\left(O_{i_1} O_{i_2}O_{i_4}O_{i_3}\right)\\
        &\text{Tr}\left(O_{i_1} O_{i_3}O_{i_2} O_{i_4}\right)+\text{Tr}\left(O_{i_1} O_{i_3}O_{i_4}O_{i_2}\right)+\text{Tr}\left(O_{i_1} O_{i_4} O_{i_2} O_{i_3} \right)+\text{Tr}\left(O_{i_1} O_{i_4} O_{i_3} O_{i_2} \right))]\\
        &\text{Tr}\left(\rho_{\alpha_1} \rho_{\alpha_2} \rho_{\alpha_3}\right)\left(\text{Tr}\left[O_{i_2} O_{i_3}O_{i_1}\right]+\text{Tr}\left(O_{i_3} O_{i_2} O_{i_1}\right)\right) \text{Tr}\left(O_{i_4}\right)\\
        &\text{Tr}\left( \rho_{\alpha_1} \rho_{\alpha_2} \rho_{\alpha_4}\right)
        \left(\text{Tr}\left(O_{i_2} O_{i_4} O_{i_1}\right)+\text{Tr}\left(O_{i_4} O_{i_2}O_{i_1}\right)\right)
        \text{Tr}\left(O_{i_3}\right)\\ 
        &\text{Tr}\left(\rho_{\alpha_1} \rho_{\alpha_3} \rho_{\alpha_4}\right)\left(\text{Tr}\left(O_{i_3} O_{i_4} O_{i_1}\right)+\text{Tr}\left(O_{i_4} O_{i_3}O_{i_1}\right)\right)\text{Tr}\left(O_{i_2}\right)\\
        &\text{Tr}\left(\rho_{\alpha_2} \rho_{\alpha_3} \rho_{\alpha_4}\right)\text{Tr}\left(O_{i_1}\right)\left(\text{Tr}\left(O_{i_3} O_{i_4} O_{i_2}\right)+\text{Tr}\left(O_{i_4} O_{i_3} O_{i_2}\right)-2 \text{Tr}\left(O_{i_2}\right) \text{Tr}\left(O_{i_3}\right) \text{Tr}\left(O_{i_4}\right)\right)]\\
        &+\mathcal{O}(\frac{1}{d^5})
    \end{split}
\end{align}

\end{widetext}

\end{document}